\documentclass[longauth]{aa}  

\usepackage{graphicx}
\usepackage{txfonts}
\usepackage{booktabs}
\usepackage{booktabs}
\usepackage{amsmath}
\usepackage{subcaption}
\usepackage{xcolor}
\usepackage[citecolor=blue, colorlinks=True]{hyperref}
\usepackage{placeins}
\usepackage{multirow}
\DeclareUnicodeCharacter{2212}{\ensuremath{-}}
\definecolor{darkgreen}{rgb}{0.0, 0.5, 0.0}

\begin{document}

\title{Looking into the faintEst WIth MUSE (LEWIS): on the nature of ultra-diffuse galaxies in the Hydra\,I cluster}
\subtitle{V. Integrated stellar population properties}

 \author{Goran Doll\inst{1, 2}\fnmsep\thanks{\email{goran.dollcarriel@inaf.it} }
          \and Chiara Buttitta\inst{1}
          \and Enrichetta Iodice\inst{1}
          \and Anna Ferr{\'e}-Mateu \inst{3,4}
          \and Jesus Falc{\'o}n-Barroso \inst{3,4}
          \and Ignacio Martín-Navarro \inst{3,4}
          \and Maurizio Paolillo \inst{1,2}
          \and Luca Rossi \inst{1,2}
          \and Duncan A. Forbes \inst{5}          
          \and Chiara Spiniello \inst{6, 1}
          \and Johanna Hartke \inst{7,8,9}
          \and Marco Gullieuszik \inst{10}
          \and Magda Arnaboldi \inst{11}
          \and Michele Cantiello\inst{12}
          \and Enrico Maria Corsini\inst{10, 13}
          \and Giuseppe D'Ago \inst{14}
          \and Michael Hilker\inst{11}
          \and Antonio La Marca\inst{15,16}
          \and Steffen Mieske \inst{17}  
          \and Marco Mirabile \inst{12,18}
          \and Marina Rejkuba \inst{11}
          \and Marilena Spavone \inst{1}
        }

        \institute{INAF $-$ Astronomical Observatory of Capodimonte, Salita Moiariello 16, I-80131, Naples, Italy
        \and
        University of Naples ``Federico II'', C.U. Monte Sant'Angelo, Via Cinthia, 80126, Naples, Italy
        \and
        Instituto de Astrofísica de Canarias, Calle V\'ia L\'actea s/n, 38200 La Laguna, Spain
        \and
        Departamento de Astrof\'isica, Universidad de La Laguna, E-38200, La Laguna, Tenerife, Spain
        \and
        Centre for Astrophysics \& Supercomputing, Swinburne University of Technology, Hawthorn VIC 3122, Australia
        \and
        Sub-Dep. of Astrophysics, Dep. of Physics, University of Oxford, Denys Wilkinson Building, Keble Road, Oxford OX1 3RH, United Kingdom
        \and
        Finnish Centre for Astronomy with ESO, (FINCA), University of Turku, FI-20014 Turku, Finland
        \and
        Tuorla Observatory, Department of Physics and Astronomy, University of Turku, FI-20014 Turku, Finland            
        \and
        Turku Collegium for Science, Medicine and Technology (TCSMT), University of Turku, FI-20014 Turku, Finland
        \and
        INAF $-$ Osservatorio Astronomico di Padova, Vicolo dell’Osservatorio 5, I-35122 Padova, Italy
        \and
        European Southern Observatory, Karl$-$Schwarzschild-Strasse 2, 85748 Garching bei München, Germany
        \and
        INAF $-$ Astronomical Observatory of Abruzzo, Via Maggini, 64100, Teramo, Italy
        \and
        Dipartimento di Fisica e Astronomia ``G. Galilei'', Universit\`a di Padova, vicolo dell'Osservatorio 3, I-35122 Padova, Italy
        \and
        Institute of Astronomy, University of Cambridge, Madingley Road, Cambridge CB3 0HA, UK 
        \and
        SRON Netherlands Institute for Space Research, Landleven 12, 9747 AD Groningen, The Netherlands
        \and
        Kapteyn Astronomical Institute, University of Groningen, Postbus 800, 9700 AV Groningen, The Netherlands
        \and
        European Southern Observatory, Alonso de Cordova 3107, Vitacura, Santiago, Chile
        \and
        Gran Sasso Science Institute, viale Francesco Crispi 7, I-67100 L'Aquila, Italy
        }

   \date{Received MM DD, YYYY; accepted MM DD, YYYY}

 
  \abstract
   {This paper presents new results from the ESO Large Programme Looking into the faintEst WIth MUSE (LEWIS).   The LEWIS sample consists of low-surface brightness galaxies (LSB) and ultra-diffuse galaxies (UDGs) located inside $0.4R_{vir}$ of the Hydra~I cluster. Integral-field spectroscopy is acquired for 24 galaxies with the MUSE spectrograph mounted on the Very Large Telescope (VLT).}
   {Our main objectives are to analyse possible correlations between the environment and the integrated stellar population properties of our targets, based on which we infer clues about their formation.}
   {For each galaxy in the sample, we extract the 1D stacked spectrum in an aperture of one effective radius $R_e$ and adopt previously published stellar kinematics to derive age, metallicity and [Mg/Fe] through full spectral fitting technique.}
   {We find that the analysed LEWIS sample has a mean metallicity of $\langle{\rm [M/H]}\rangle=-0.9\pm0.2$~dex and a mean age of $10\pm 2$~Gyr, comparable to previous results of UDGs in other clusters. 
   According to the position in the projected phase space, galaxies can be classified into two groups: early infallers galaxies, which on average have slightly higher metallicities ($\langle{\rm [M/H]}\rangle_{\rm early} = -0.8~\pm~0.1$\,dex), and late infallers galaxies, with slightly lower values ($\langle{\rm [M/H]}\rangle_{\rm late} = -1.0~\pm~0.1$\,dex). 
   {  According to their properties, late-infallers tend to be rotation-supported systems. Conversely, two types of galaxies are found in the early-infall region. Roughly half have metallicities consistent with the dwarf galaxy mass–metallicity relation. The other half show higher metallicities (with $\langle{\rm [M/H]}\rangle \geq -1.0$\,dex), and are located outside the $1\sigma$ scatter of the mass-metallicity relation. The two sub-groups of early-infallers also display different timescales for stellar mass assembly. Metal-rich galaxies reached 50\% of their stellar mass in less than 1~Gyr and show a prolonged and almost constant star formation over more than 12 Gyrs. 
   The other galaxies exhibit a star-formation history similar to that found for galaxies in the late-infall region. Both early and late infallers show solar-like $\alpha$ abundances.}}
   {  {From the analysis of stellar population properties presented in this work and stellar kinematics previously obtained from LEWIS, we have identified different classes of UDGs within the Hydra~I cluster   {as shown by metallicities, quenching timescales and kinematics}, which   {suggest} different formation mechanisms taking place. Almost all of the UDGs and LSBs in this cluster are consistent with the puffed-up dwarf formation scenario, having dwarf-like metallicities and being consistent with the mass-metallicity relation for dwarfs. In the innermost regions of the cluster, where more metal-rich UDGs are found, tidal effects or the environment might have influenced their formation and evolution.}}


\titlerunning{Stellar populations properties}
\authorrunning{G. Doll et al.}
\maketitle
%

\section{Introduction}
\label{sec:intro}

Ultra-diffuse galaxies (UDGs) are a distinct and intriguing class of low-surface-brightness (LSB) galaxies characterized by extremely low surface brightness ($\mu_{0,g}> 24$\,mag\,arcsec$^{-2}$), large effective radii ($R_{\rm e}>1.5$\,kpc), and small stellar masses \citep[$M \sim 10^{7} - 10^{8.5} M_{\odot} $, see][for a review on UDG definition]{VanNest2022}. Originally detected in the 1980s (e.g., \citealp{Sandage1984}), these galaxies were rediscovered in large numbers in 2015 in the Coma cluster, thanks to advances in observational techniques and the development of instrumentation capable of probing LSB galaxies (\citealp{vanDokkum2015}; \citealp{Koda2015}). Since then, UDGs have been found in a wide range of environments, from galaxy clusters (e.g., \citealp{Mihos2015}; \citealp{Mancera-pina2018udg}; \citealp{Lim2020}; \citealp{LaMarca2022b}) to groups (e.g., \citealp{Trujillo2017}; \citealp{Forbes2019}) and field (e.g., \citealp{MartinezDelgado2016}; \citealp{Bellazzini2017}). 

Many observational studies have shown that UDGs span a wide range of properties in relation to the environment where they reside. For instance, UDGs in dense environments are found to be red, quenched, and spheroidal \citep{Koda2015, vanDokkum2015, vanDokkum2016, RuizLara2018, Lee2020}. On the other hand, UDGs in less dense environments, such as in the field or group outskirts, are often bluer, irregular, and gas-rich \citep{VanderBurg2016, Leisman2017, Roman2017a, Prole2019a}. UDGs have even been identified at redshift $z \sim 0.3$, providing evidence for their existence at earlier cosmic times \citep{Janssens2017}.

Theories regarding the formation of UDGs are diverse, with several mechanisms proposed to explain their distinct properties. A leading scenario is that UDGs are simply the extreme end of the distribution of dwarf galaxies, ``puffed-up'' by either internal or external processes, which lead to large sizes and low surface brightness. Internal mechanisms include stellar feedback and/or high spins of the dark matter halo \citep{Amorisco2016, diCintio2017, Rong2017, Tremmel2020}. In contrast, models invoking external environmental effects suggest that tidal forces or ram-pressure stripping from a cluster environment could strip gas and quench star formation, transforming a typical dwarf galaxy into a passive, low-surface-brightness object \citep{Yozin2015, Ogiya2018, Bennet2018, Muller2019, Poggianti2019, Benavides2023}.

Another suggested formation model comes from the ``failed-galaxy'' hypothesis, where UDGs are remnants of early, dark matter-dominated systems that were quenched very early in cosmic history \citep{vanDokkum2015, Peng2016}. These ``failed'' galaxies are hypothesised to have been incapable of accreting sufficient gas or sustaining star formation, leaving them with small stellar masses but residing in large, dark matter-dominated halos. While this model has been difficult to reproduce in cosmological simulations, it could explain the massive dark matter halos observed in some UDGs \citep{Beasley2016, Gannon2022, Forbes2024}.

Simulations have provided a major advance in our understanding of UDGs. For instance, \cite{Sales2020} used the Illustris TNG100-1 simulation to study the formation of UDGs, distinguishing between two main populations: UDGs that are ``born'' diffuse (B-UDGs) and UDGs that are transformed into diffuse galaxies through tidal stripping (T-UDGs). B-UDGs are typically field galaxies that retain their low surface brightness even after being accreted into clusters, while T-UDGs have experienced significant mass loss due to tidal interactions after entering the cluster environment. According to \cite{Sales2020}, T-UDGs are older, more metal-rich, and have lower velocity dispersions than B-UDGs. The distinction between these two populations, which has been supported by additional observational work \citep[e.g.,][]{Gannon2020, Ferre-Mateu2018, FerreMateu2023}, offers a powerful framework for understanding the diverse properties of UDGs across different environments.

Large-scale photometric surveys like the Systematically Measuring Ultra Diffuse Galaxies survey \citep[SMUDGes, ][]{Zaritsky2023} provide a broad overview of the UDG population but have limitations in directly comparing stellar populations - particularly in terms of mass and metallicity - due to inconsistencies in distance measurements and the lack of direct metallicity data. Recently, \cite{Buzzo2024b} has provided the most comprehensive analysis of UDGs' properties by systematically re-examining UDG samples from \cite{Buzzo2022} and \cite{Buzzo2024} alongside a new nearly ultra-diffuse galaxy (NUDGes) sample, creating a large, unified dataset for statistical studies. The authors have identified two distinct subclasses of UDGs: Class A UDGs, which are less massive, elongated, and found in low-density environments, and Class B UDGs, which are more massive, rounder, and located in denser environments. These results suggest that Class A UDGs align with the puffed-up dwarf scenario, where internal processes such as stellar feedback and gas expulsion drive their expansion, whereas Class B UDGs may be better explained by the failed galaxy formation scenario, possibly involving early quenching in dense environments. 

Spectroscopic studies of UDGs have provided further constraints on stellar population properties of UDGs. Those formed through internal feedback processes tend to have extended star formation histories (SFHs) and younger ages ($\sim6-9$~Gyr), with metallicities comparable to classical dwarfs ([M/H]$\sim-0.8$\,dex). In contrast, UDGs formed via the failed galaxy mechanism have older stellar populations ($\sim10-12$~Gyr) and much lower metallicities ([M/H]$\sim -1.5$\,dex; see e.g., \citealp{Ferre-Mateu2018,FerreMateu2023, RuizLara2018, Levitskiy2025arXiv}). Studying the largest sample of UDGs with spectroscopic data until LEWIS, and covering different environments and globular cluster (GC) richness, \citet{FerreMateu2023} found that the local environment is relevant for the formation pathway of UDGs. In high-density environments, UDGs seem to have formed stars very early and have short quenching times ($\sim6$ Gyr). Conversely, UDGs in low-density environments (as groups and cluster outskirts), have younger ages and a prolonged SFH. These results are compatible with the inferred [$\alpha$/Fe] ratios of UDGs, which are overall elevated but particularly for the high-density environment ($\ge 0.3$\,dex). 

In summary, having a large statistical spectroscopic sample is fundamental to study not only the stellar population properties, but also other structural properties such as the amount of dark matter (DM), the stellar kinematics, and the stellar population properties of globular cluster content. In this paper, we use spectroscopic data from LEWIS\footnote{\href{https://sites.google.com/inaf.it/lewis/home}{https://sites.google.com/inaf.it/lewis/home}}. The Looking into the faintEst WIth MUSE (P.I. E. Iodice) is an ESO Large Programme approved in 2021, with 133.5 hours granted with the Multi Unit Spectroscopic Explorer (MUSE@VLT). LEWIS data have mapped, for the first time, a sample of 30 LSB galaxies in the Hydra-I cluster. These galaxies represent the most extreme tail of the surface-brightness versus size distribution \citep{Iodice2020, LaMarca2022b} mapped in Hydra-I from the VEGAS survey\footnote{\href{https://sites.google.com/inaf.it/vegas/home}{https://sites.google.com/inaf.it/vegas/home}}. The majority (22) of LEWIS galaxies are classified as UDGs according to the \citealp{vanDokkum2015} definition. The rest are extreme LSBs, close to this definition but slightly brighter and/or smaller, akin to the NUDGEs (nearly-UDGs) definition from \citealp{Forbes2024}. Unlike most previous spectroscopic studies of UDGs, which compiled samples of galaxies from various environments, the LEWIS sample is homogeneous, allowing for the study of a substantial number of galaxies within a single environment. This is crucial for understanding how the environment affects galaxies' properties and discriminating between different classes of UDGs. 

{The LEWIS project has been extensively presented in \citealt{Iodice2023} (hereafter Paper~I) together with the analysis of the location of LEWIS galaxies in the projected phase-space \citep{Forbes2023}. The LEWIS data allowed us to derive, for the first time, the integrated and spatially resolved stellar kinematics for a substantial sample of UDGs \citep[][hereafter Paper~II]{Buttitta2025}. In LEWIS Paper~III \citep{Hartke2025}, we constrained the formation history of the intriguing UDG32, finding that it has been formed from pre-processed material (see also \citealp[]{Iodice2021}) and in Paper~IV (Mirabile et al. submitted), we constrain the characteristics of globular cluster populations of four UDGs.}

In this paper, we focus on the integrated stellar population properties of galaxies in the LEWIS sample. The Paper is organised as follows. In Section~\ref{sec:data_obs} we briefly present the LEWIS data. In Section~\ref{sec:methods}, we describe the methodology adopted to derive the stellar population properties and star-formation histories. In Section~\ref{sec:results}, we present the properties of the LEWIS sample, while in Section~\ref{sec:discussion}, we discuss the plausible formation scenarios for the LEWIS UDGs.   {Finally, in Section~\ref{sec:conclusion}, we report the conclusions and present the future perspectives.}

The Hydra I cluster (Abell 1060, $\alpha\ {\rm (J2000)}={\rm 10h\,36m\,42.7100s} , \delta\ ({\rm J2000}) = {\rm -27d\,31m\,42.900s}$) is located a distance of $51\pm6$\,Mpc \citep{Christlein2003}, such that $1'' = 247.3\,{\rm pc}$.

\section{Data}
\label{sec:data_obs}

\subsection{LEWIS sample: observations and data reduction}
\label{subsec:obs}

Observations for the LEWIS survey were conducted using the MUSE integral-field spectrograph \citep{Bacon2010} at the European Southern Observatory (ESO, Program Id. 108.222P, P.I. Iodice). MUSE was set to operate in Wide Field Mode without AO, offering a field-of-view (FOV) of 1'$\times$1' and a spatial resolution of 0.2~arcseconds per pixel. The data covers a wavelength range of 4800-9300~\AA\ with a spectral sampling of 1.25~\AA/pixel and a mostly flat resolution profile (FWHM~=~2.69~\AA\,; see Paper~II for the details on the instrumental line-spread function). 

The LEWIS sample presented in Paper~I is composed of 30 galaxies, 22 of them classified as UDGs and eight are LSBs. From these 30 galaxies of the original sample, 5 were excluded because of the presence of Ferris wheel-like patterns,  i.e. the light diffraction of a bright star outside the field of view, or because the exposure was heavily contaminated by the presence of a bright, saturated star or a nearby galaxy. 
  
From the entire LEWIS sample of 30 galaxies, four UDGs had partial observations at the time this analysis was conducted. We therefore considered only the 26 remaining galaxies (see Table~2 from Paper II).
Of these, UDG32 was further excluded because it was already the topic of a dedicated paper \citep{Hartke2025}.
In Section~\ref{sec:results}, we further discuss the quality criteria we adopted to redefine the final sample for which accurate measures of stellar population properties were derived. The observations for LEWIS were completed in March 2025. The observations and data reduction of the final LEWIS sample are described in detail in Paper~II.

\section{Analysis}
\label{sec:methods}

\subsection{Age and metallicity}
\label{subsec:methods:age_met}

To obtain the mean stellar population properties of the LEWIS galaxies, we fit the stacked spectrum extracted from an aperture of radius equal to an effective radius 1\,$R_{\rm e}$  with the package {\tt Penalized Pixel Fitting} \citep[{\tt ppxf},][]{Cappellari2017, Cappellari2023}. The code uses single stellar population synthesis (SSP) models and allows masking of gas emission lines and low-S/N regions. Moreover, it allows multiplicative and/or additive polynomials to correct for the shape of the continuum.

As templates, we use the entire set of SSP models from sMILES \citep{Knowles23}, with BaSTI isochrones, assuming a Kroupa initial mass function with a fixed slope of 1.3. The metallicities span from $-1.79$ to $+0.26$\,dex, from 0.03 to 14 Gyr in age, and from $-0.2$ to $+0.6$ in [$\alpha/$Fe]. The initial setup for the fitting is based on the main results from Paper II:

\begin{itemize}
    \item {Wavelength range}: we use the narrow region between 4800\,\AA\,-\,5400\,\AA. This is a relevant spectral range for stellar population studies, since it contains age- and metallicity-sensitive absorption lines, particularly for old stellar populations. This is also the spectral range where we have the best quality of the MUSE data.
    \item {Instrumental   {line spread function (LSF)}}: we adopt the LSF measured from Paper~II. Given that our LSF is relatively flat, we fix the template resolution at 2.69 $\AA$ for ease and speed of iterative computations.
    \item {Polynomial order:} The literature frequently uses multiplicative and additive polynomials to account for spectral calibration issues and template mismatching (see e.g, \citealp{Cappellari2017}). To avoid potential biases for the computation of the stellar population properties, we decided to use only multiplicative polynomials of order 6 to 10, using Paper II as a reference point.
\end{itemize}

The output of {\tt ppxf} is a grid of weights assigned to each SSP model of the bestfit solution that allows obtaining mass or light-weighted age and metallicity of the observed galaxy's spectrum. We run a first fitting iteration at the 4800 \AA -- 5400 \AA \, on the rest-frame spectra, masking consistently noisy regions and skyline residuals. From residuals of the first iteration, we set a masking threshold of $3\sigma$ that determines the number of pixels to be masked in a second iteration by using a simple sigma-clipping technique. To estimate the uncertainties in the stellar population properties, we follow an approach similar to the one described in Paper~II: we implement a bootstrap-like procedure to generate several perturbations of the spectrum by adding Poissonian noise, which introduces a standard deviation in the S/N per pixel distribution of less than 3$\sigma$. Then we fit the perturbed spectra by exploring a grid of randomly generated input parameters, emulating standard Monte Carlo approaches. We derive the mass-weighted metallicities and mean ages for the entire LEWIS stellar population sample, which are reported in Table~\ref{tab:stellar_pop}. 

\subsection{Star formation histories}
\label{subsec:methods:SFHs}

The analysis of the SFHs can reveal detailed evolutionary patterns of galaxies, offering a richer and more comprehensive understanding of their formation and evolution. Star formation (SF) quantities require a precise reconstruction of the SFH of the observed spectrum. From the output of {\tt ppxf}it is possible to construct the SFH by collapsing the weights along the metallicity axis to generate a cumulative mass function that describes how the mass of a galaxy is built across cosmic time. Alternatively, one can construct the SFH by dividing the specific mass fraction generated by each SF episode by its associated timeframe or age bin. We follow the first approach to construct cumulative mass fraction plots and derive age percentiles for the formed stellar mass. 

From the recovered SFHs, we calculate two characteristic \textit{lookback times:} when the galaxies formed $50\%$ ($t_{50}$) and $90\%$ ($t_{90}$) of their stellar mass, respectively. The value $t_{90}$ is often used as a proxy for the quenching time, as the point at which the SF has effectively stopped  \citep{FerreMateu2023}.   {While most of our galaxies tend to present a single star formation episode at early epochs, a few exceptions do show a shorter, minor recent starburst. For this reason, we also commonly use $t_{50}$ as a proxy for quenching throughout the paper, as it tends to be better representative of the former, oldest star-formation episode shutdown.} To quantify the evolutionary phase, we define the duration of the \textit{ formation phase}, which measures the time elapsed between two stages of SF and allows us to separate the evolution of the galaxy into two phases: the first phase $\Delta t_{50} = t_{\rm BB} - t_{\rm 50} $ where we set $t_{\rm BB} = 14$\,Gyr, and the second phase $\Delta t_{90} = t_{50} - t_{90}$ (see \citealt{GrebolTomas2023}). In the first phase, we neglect the time that passed since the Big Bang and the onset of the first episode of SF in our galaxies, as adopted in \citet{FerreMateu2023}, because LEWIS galaxies tend to be old. Lower values of $\Delta t$ mean a rapid SF, while higher values suggest that SF proceeds more slowly.

Given that different sets of weights can replicate the observed spectrum, these inherent degeneracies are commonly addressed through a regularisation technique, which selects the smoothest solution, suppressing solutions that exhibit abrupt variations in the age-metallicity plane \citep{Cappellari2017}. While this introduces a potential bias, it typically generates a more realistic model. We test the effects of regularisation and find no significant difference in the median recovered metallicity and age, similarly to what was reported previously in \citet{FerreMateu2023}. Thus, we decide to use the non-regularised solutions: we expect these to be comparable to literature SFHs of other UDGs, despite having a slightly `burstier' shape. It is important to stress that this choice, combined with the low S/N of some sources, means that the recovered SFH for any individual galaxy can be sensitive to noise. Considering that these quantities can be affected by the regularisation, we focus exclusively on the median trends of SFH rather than considering individual SFH to identify the broad differences in their evolutionary pathways.

From the mass-weighted metallicity weights, we derived the mass-to-light ($M/L$) ratio for each galaxy, using the corresponding transmission curves to compute photometric measurements for our spectra. Subsequently, we also get spectroscopic stellar masses ($M_\ast$) by combining $M/L$ ratios with the photometric $r$-band luminosity (Paper~I). To construct a normalised SFH, we divide each collapsed mass weight by the age width of the corresponding grid template. Appendix~\ref{appendix} presents, for each galaxy, all the relevant stellar population information, including their {\tt ppxf} fit, resulting age-metallicity weights, and their derived SFH.

\begin{table*}[!h]
\setlength{\tabcolsep}{3.5pt}
\renewcommand{\arraystretch}{1.1}
\centering
\caption{Stellar populations properties of UDGs and LSBs in Hydra\,I.}
\label{tab:stellar_pop}
\begin{tabular}{lcccccccc}
\hline
\hline
Object & Position in Hydra &   Age  &  [M/H] &       $M_\ast$     & [Mg/Fe] & [$\alpha$/Fe] & $t_{90}$ & $t_{50}$  \\       
       &                   &  [Gyr] &  [dex] & [10$^8$ M$_\odot$] &  [dex]  &  [dex]        & [Gyr]    &  [Gyr]    \\
(1)    & (2)               &   (3)  &  (4)   &         (5)        &    (6)  &     (7)       &  (8)     &    (9)    \\
\hline
UDG 1  & Very early infall &  $7.6  \pm 2.9$ & $-0.5 \pm 0.1$ & $1.73 \pm 0.43$ & $-0.8 \pm 0.2$ & $-0.4 \pm 0.1$ &  $1.0 \pm 2.8$ & $13.6 \pm 7.5$ \\
UDG 4  & Late infall       &  $13.7 \pm 0.2$ & $-0.9 \pm 0.1$ & $9.07 \pm 0.28$ & $0.3 \pm 0.2$  & $0.2 \pm 0.1$  & $13.3 \pm 0.1$ & $13.7 \pm 0.1$ \\
UDG 8  & Late infall       &  $8.4  \pm 0.7$ & $-1.2 \pm 0.1$ & $0.71 \pm 0.04$ & $0.1 \pm 0.2$  & $0.1 \pm 0.1$  &  $8.6 \pm 0.6$ & $9.2 \pm 0.7$  \\
UDG 9  & Very early infall &  $12.7 \pm 0.6$ & $-0.7 \pm 0.1$ & $2.62 \pm 0.22$ & $1.1 \pm 0.4$  & $0.6 \pm 0.2$  & $13.1 \pm 0.2$ & $13.3 \pm 0.5$ \\
UDG 11 & Very early infall &  $7.3  \pm 0.3$ & $-1.3 \pm 0.1$ & $1.41 \pm 0.05$ & $0.4 \pm 0.2$  & $0.2 \pm 0.1$  &  $6.9 \pm 0.4$ & $7.8 \pm 0.4$  \\
UDG 12 & Very early infall &  $9.0  \pm 0.5$ & $-0.9 \pm 0.1$ & $1.25 \pm 0.07$ & $0.3 \pm 0.2$  & $0.2 \pm 0.1$  &  $7.6 \pm 1.4$ & $8.0 \pm 1.0$  \\
\hline
\hline
LSB 5  & Very early infall &  $8.6  \pm 1.4$ & $-0.7 \pm 0.2$ & $0.85 \pm 0.12$ & $-1.1 \pm 0.8$ & $-0.6 \pm 0.4$ & $0.8 \pm 0.2$ & $13.6 \pm 0.6$ \\
LSB 6  & Late infall       &  $8.9  \pm 0.7$ & $-1.0 \pm 0.1$ & $3.55 \pm 0.24$ & $-0.1 \pm 0.2$ & $0.0 \pm 0.1$  & $7.7 \pm 1.6$ &  $9.0 \pm 0.7$ \\
LSB 7  & Very early infall &  $9.2  \pm 1.0$ & $-0.7 \pm 0.1$ & $4.39 \pm 0.41$ & $-0.1 \pm 0.3$ & $0.0 \pm 0.2$  & $9.3 \pm 0.7$ & $11.1 \pm 1.1$ \\
LSB 8  & Late infall       &  $11.1 \pm 1.0$ & $-0.8 \pm 0.1$ & $3.31 \pm 0.24$ & $-0.7 \pm 0.3$ & $-0.4 \pm 0.2$ & $9.5 \pm 0.9$ & $11.1 \pm 1.4$ \\
\hline
\bottomrule
\end{tabular}
\tablefoot{Column 1 reports the target name in the LEWIS sample. Column 2 reports the position of the LEWIS sample in Hydra\,I, according to the infall diagnostic diagram. Columns 3 to 6 report the age, metallicity, stellar masses and alpha-enhancements derived from stellar population analysis. Column 7 reports the derived [$\alpha$/Fe] computed using the formula [$\alpha$/Fe] = 0.02 + 0.56 $\times$ [Mg/Fe]. Columns 8 and 9 report the 90th and 50th percentiles of star formation times, respectively.}
\end{table*}

\subsection{Chemical abundances}
\label{subsec:methods:abundances}

The chemical abundance ratios provide another independent constraint on the star formation timescale. For example, unlike the SFH lookback times, which reflect the totality of the mass assembly, the [$\alpha$/Fe] ratio is a chemical clock for the phase and efficiency of the initial burst of SF. A high ratio is effectively the blueprint of a very short and intense timescale ($< 1$ Gyr), where enrichment from Type-II supernovae (producing heavy elements such as Mg) massively outpaces the slow enrichment from Type-Ia supernovae \citep{Worthey1992}.

To characterise abundances in our sample, we measure iron and magnesium indices, along with the H$\beta_0$ as an age indicator. Due to our wavelength selection and median S/N, we decide to work exclusively with the [Mg/Fe] abundance as a [$\alpha$/Fe] tracer, similarly to what has been done in the literature. The line index measurement is done with the latest version of {\tt pyphot}\footnote{\href{https://mfouesneau.github.io/pyphot/}{https://mfouesneau.github.io/pyphot/}}, and uncertainties are measured from perturbed spectra (injecting Poisson noise into the raw spectrum), also utilised for other error estimations. We adopt a combination of the following two methods:

\begin{enumerate}
    \item We use the sMILES templates to construct model grids of Mg as a function of Fe (Fig.~\ref{fig.Licks-MgFe}). We use a regular, arithmetic average of Fe5015, Fe5270 and Fe5335, compared to the Mg$_{b}$ index, for different values of alpha enhancement ([$-$0.2, $+$0.0, $+$0.2, $+$0.4, $+$0.6] dex), convolved to the spectral resolution of our data. In age and metallicity, we centre these grids at the {\tt ppxf}-derived values. We interpolate according to the position of the data within the model grids.
    \item Using the scaled-solar E-MILES models \citep{Vazdekis2016}, we construct model grids of H$\beta_0$ versus different metallicity ones, mainly Fe and Mg, to estimate the Mg and Fe enhancement on each galaxy from the proxy [Mg/Fe]\,$\,= [Z_{\rm Mg}/Z_{\rm Fe}] = Z_{\rm Mg}-Z_{\rm Fe}$ \citep{Vazdekis2015}. However, due to the   {measured chemical composition of these galaxies}, which makes many LEWIS measurements fall outside the range of the model grids (see Fig.~\ref{fig.Licks-age}, upper panel) or in a very degenerate region, we replace the H$\beta_0$ indices by the proper light-weighted median age \citep{Ferre-Mateu2014} from {\tt ppxf}. This produces a more orthogonal model grid facilitating interpolation (see Fig.~\ref{fig.Licks-age}, bottom panel)
\end{enumerate}

The associated uncertainties for the derived [Mg/Fe] are obtained from Monte Carlo simulations of the different line indices involved. Overall, we find that the estimates from both methods are consistent; hence, we use the average of both for the final [Mg/Fe] value. The only exceptions are UDG\,1 and LSB\,5, for which the second method did not provide reliable results, and thus we limit them to the results of method 1. 

We estimate the [$\alpha$/Fe] abundance using the formula:

\begin{equation}
   \label{eq:alpha}
   [\alpha/{\rm Fe}]=0.02+0.56\times[Z_{\rm Mg}/Z_{\rm Fe}] 
\end{equation}

\noindent where we assume that sMILES models (with BaSTI isochrones) behave similarly to the regular Base E-MILES models, on which this formula is based \citep{Vazdekis2015}.

\section{Results: stellar populations properties of LEWIS galaxies}
\label{sec:results}

In Table~\ref{tab:stellar_pop}, we report all the quantities derived from the analysis of stellar population properties for those galaxies which have constrained kinematics \citep[see Table~2 from][for details]{Buttitta2025}. They are the ages, metallicities [M/H], stellar masses $M_\ast$, chemical abundances [Mg/Fe] and characteristic lookback times ($t_{50}$ and $t_{90}$). In addition, we report the classification in the projected phase-space (PPS) diagram as presented in Paper~II. 

The accuracy of the results strongly depends on the quality of the spectra and the reliability of stellar kinematics parameters. In particular, the effective velocity dispersion $\sigma_{\rm eff}$ could be sensibly overestimated in poor-quality spectra, and thus cannot be used in the extraction of stellar population properties (see Paper~II for details). The final LEWIS sample presented in Paper~II is composed of 25 objects (19 UDGs and 6 LSBs), but only for a subsample of them, the stellar kinematics information was accurately derived. In this work, we use the same classification presented in Paper~II to identify galaxies with precise measurements of stellar kinematics parameters and thus with reliable stellar population properties. The `C' (constrained fit) galaxies have high-quality spectra and reliable estimates of stellar kinematics; the `I' (intermediate fit) objects might have biased stellar kinematics parameters due to the presence in their spectra of sky line residuals close to absorption features, while `U' (unconstrained fit) galaxies have low quality spectra and thus the spectral fitting is unreliable.

In this work, we present accurate measurements for only constrained cases (10 galaxies), six of which are classified as UDGs, and the remaining four are LSB galaxies. The main conclusions presented in this work are based only on this sample, whereas results of the stellar population analysis of the intermediate cases (UDG3, UDG7,  UDG10, LSB1, and LSB4) are discussed in Appendix~\ref{app:intermediate}.

\begin{figure*}[!h]
\centering
\includegraphics[scale=0.44]{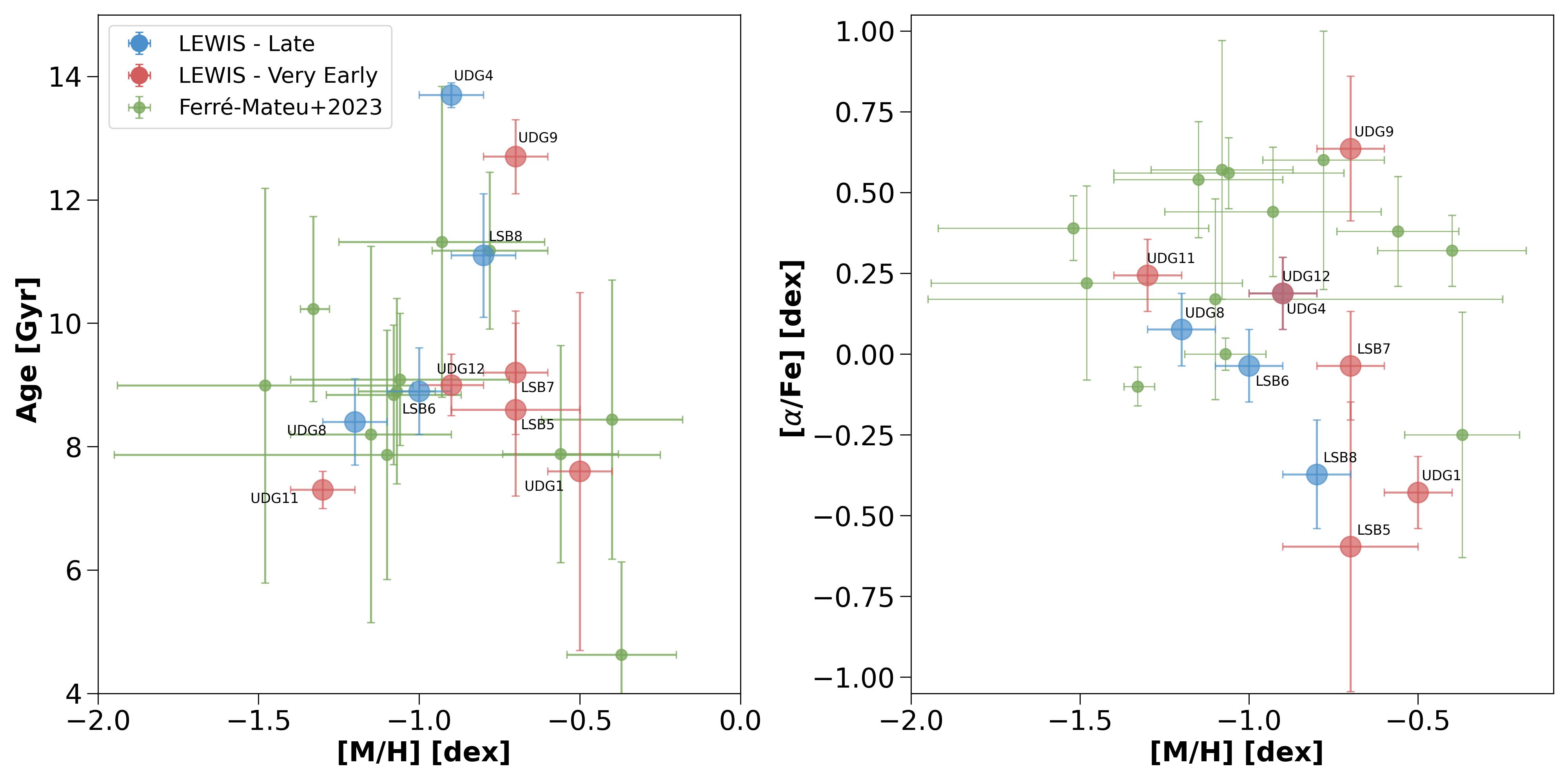}
\caption{  Stellar population properties of LEWIS galaxies. Left panel: Mean mass-weighted ages as a function of metallicities [M/H] for the LEWIS sample (coloured datapoints) and the literature UDGs (green circles, from \citealp{FerreMateu2023}). The red and blue circles represent galaxies classified as very early and late infallers, respectively. Right panel: chemical abundance [$\alpha$/Fe] as a function of metallicities [M/H]. Datapoints are colour-coded as in the left panel.}
\label{fig:ages_mets_alpha}
\end{figure*}

\subsection{Age, metallicities and chemical abundance}
\label{sec:properties}

Fig.~\ref{fig:ages_mets_alpha} (left panel) shows the mean mass-weighted ages as a function of metallicity [M/H] for the LEWIS sample and literature data on cluster UDGs from \cite{FerreMateu2023}. For LEWIS, we find a mean metallicity $\langle{\rm [M/H]}\rangle = -0.9\pm 0.2$\,dex and an average age of $\langle{\rm Age}\rangle = 10 \pm 2$\,Gyr, consistent with these literature studies ($\langle{\rm Age}\rangle = 8.9 \pm 2.7$\,Gyr and $\langle {\rm [M/H]}\rangle = -1.0 \pm 0.4$\,dex). We find that the derived ages of LEWIS UDGs are also consistent with the ages of MATLAS classical dwarfs analysed in \citealt{Heesters2023} (${\rm Age}\sim6-14$\,Gyr), although the latter are on average more metal-poor overall ([M/H] = $-1.9$ to $-1.3$\,dex)  {, coming from low-to-moderate density environments.}. The dwarf galaxy sample of the SAMI-Fornax survey \citep{Romero-Gomez2023} instead shows slightly younger ($\langle{\rm Age}\rangle\sim 8$\,Gyr) and more metal-rich ([M/H] = $-0.4$\,dex) values, but a direct comparison is hard to establish since they only report luminosity-weighted quantities. 

In Fig.~\ref{fig:ages_mets_alpha}, we marked with different colors the LEWIS galaxies belonging to the two regions of the projected phase-space (PPS), i.e. very early and late infallers (see also column 2 in Table~\ref{tab:stellar_pop}). To remind the reader, very early infallers are those galaxies that entered the cluster at least $\sim7$\,Gyr ago and thus are virialised with the cluster, whereas late infaller galaxies are approaching the cluster for the first time or have already passed the first pericenter   \citep[Fig.~\ref{fig:PPS}, see also][]{Rhee2017, Forbes2023}. This infall diagnostic diagram allows us to highlight possible dependencies between structural properties of the galaxies and the cluster. We notice that overall, early infall galaxies appear to have somewhat higher metallicities compared to late infall ones.

Fig.~\ref{fig:ages_mets_alpha} (right panel) shows the chemical abundance [$\alpha$/Fe] as a function of the metallicity [M/H] for the LEWIS sample and literature data of cluster UDGs from \cite{FerreMateu2023}. We compute the [$\alpha$/Fe] using Eq.~\eqref{eq:alpha} to directly compare our findings with UDGs in the literature. Overall, LEWIS galaxies show values similar to the UDGs in literature from \cite{FerreMateu2023}, ranging from sub-solar ([$\alpha$/Fe]$\,\sim$ -0.5\,dex) to super-solar ([$\alpha$/Fe]$\,\sim$ +0.5\,dex) abundances.  Except for UDG1, we note that all the sub-solar objects are LSBs, whereas other sources display an enhancement consistent with regular dwarfs or slightly above, as observed in the literature data for UDGs.

\begin{figure*}
\centering
\includegraphics[scale=0.72]{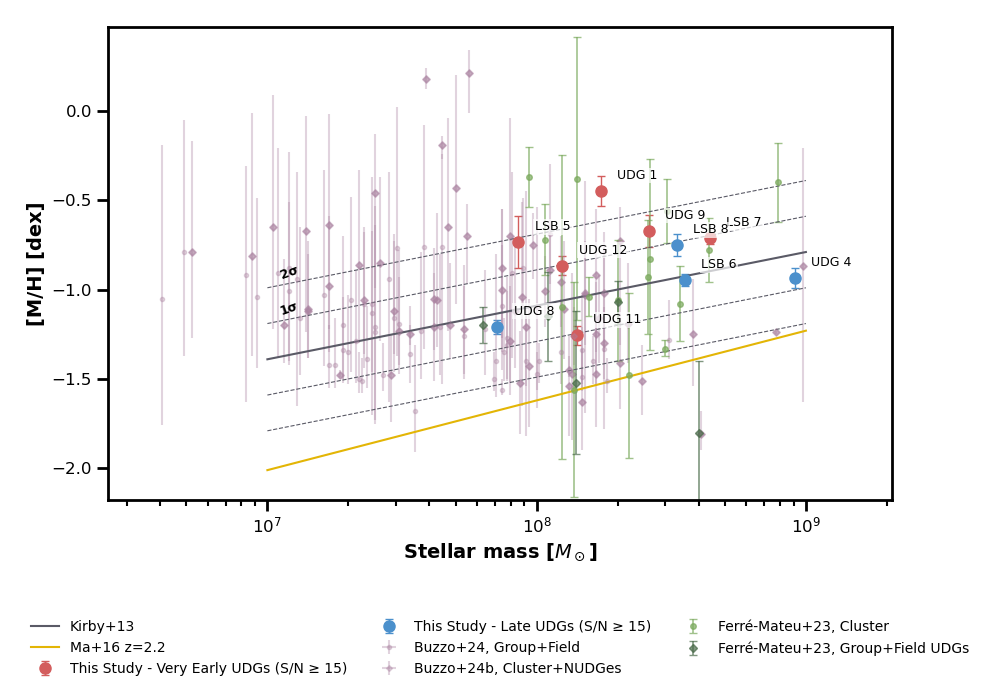}
\caption{Mass-metallicity relation. LEWIS datapoints are colour-coded as in Fig.~\ref{fig:ages_mets_alpha},   {while other coloured datapoints correspond to literature samples: UDGs in clusters (light green circles) and in low-density environments (dark green diamonds) from \citealp{FerreMateu2023}, UDGs in low-density environments (violet circles) from \citealp{Buzzo2024}, and UDGs and nearly UDGs in high-density environments (violet diamonds) from \citealp{Buzzo2024b}. The black solid line is the mass-metallicity relation found for dwarf galaxies assuming a constant $\alpha$-enhancement \citep{Kirby2013}, and the grey dashed lines represent the $1\sigma$ and $2\sigma$ confidence regions of the relation.} The yellow line represents the evolving mass-metallicity relationship at redshift $z \sim 2.2$ from \cite{Ma2016}.}
\label{fig:MZR}
\end{figure*}


\subsection{The mass-metallicity relation}
\label{sec:MZR}

Fig.~\ref{fig:MZR} shows the mass-metallicity relation for all the LEWIS galaxies analysed in this work, compared with other works in the literature that analysed samples of UDGs and dwarf galaxies in different environments. Considering the uncertainties, 7 out of 10 galaxies analysed in this work have metallicity consistent with the relation found for the dwarf galaxies \citep{Kirby2013} and with most of the UDGs studied by \citet{Buzzo2024} and \citet{FerreMateu2023}. The remaining 3 galaxies (UDG1, UDG9 and LSB5) have higher metallicities than dwarf galaxies with similar stellar mass and are located outside the $1\sigma$ confidence region of the \cite{Kirby2013} mass-metallicity relation for dwarfs. 
Values are consistent with the observed scatter for the UDGs in the literature. 

Different from the UDGs samples by \cite{Buzzo2024} and \cite{FerreMateu2023}, at stellar masses larger than $10^8\,M_\odot$, we do not find LEWIS galaxies having metallicities lower than those expected 
for dwarf galaxies of comparable stellar masses, i.e. ${\rm [M/H]} \lesssim −1.5$\,dex. 
This result can have implications for the formation channels for UDGs and will be discussed in Sec.~\ref{sec:discussion}.

Finally, according to the PPS, we notice that late-infallers overall have metallicities consistent with the \cite{Kirby2013} relation for dwarfs, whereas early-infallers are either consistent or present higher values.

\subsection{Star formation histories and quenching times}

\begin{figure*}[t]
\centering
\includegraphics[scale=0.45]{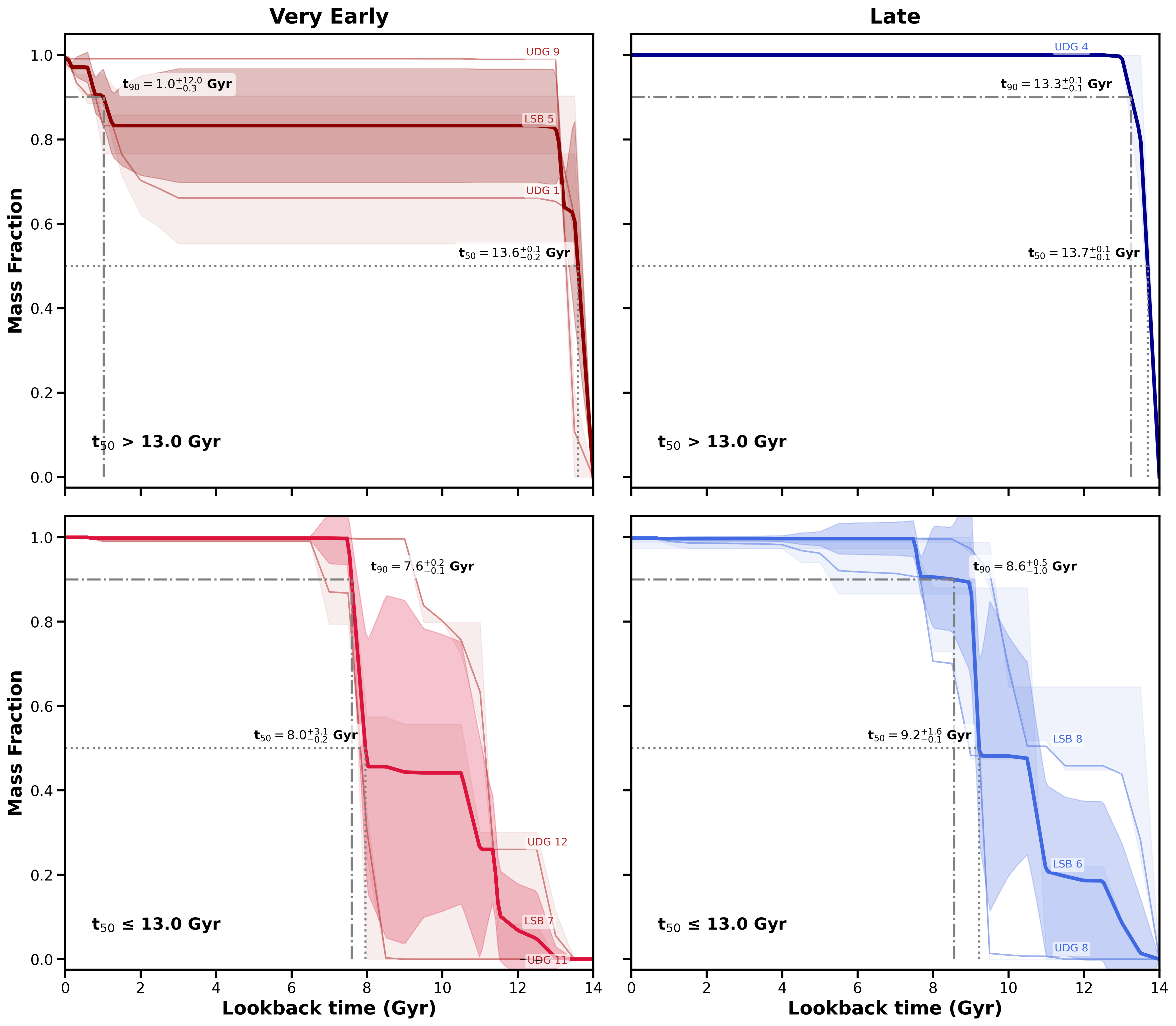}
    \caption{Star-formation histories of LEWIS galaxies, where we have separated very early quenching ($t_{50} > 13$ Gyr, upper panels) from normal to late quenching ($t_{50} < 13$ Gyr, lower panels). Lines are colour-coded as in Fig.~\ref{fig:ages_mets_alpha}, separating very early and late infallers. In each panel, the thin lines represent the individual SFHs of galaxies, whereas the thick solid line represents the median SFH. Grey lines represent the percentiles $t_{50}$ and $t_{90}$ of the mean SFHs.}
    \label{fig:SFH_alpha}
\end{figure*}

Fig.~\ref{fig:SFH_alpha} shows the star-formation histories (SFH) of LEWIS galaxies  {, separated by their PPS positioning and quenching times as traced by $\Delta t_{50}$}. For the late infaller group (right panels), we find a median lookback time of $t_{90} = 8.6^{+0.5}_{-1.0}$\,Gyr and $t_{50} = 9.2^{+1.6}_{-0.1}$\,Gyr. The above estimates exclude UDG4, which shows a different SFH (see top-right panel of Fig.~\ref{fig:SFH_alpha}). This object quenched earlier ($t_{90} = 13.3^{+0.1}_{-0.1}$\,Gyr) and very fast ($\Delta t_{90} = \sim 0.7$\,Gyr).

We find that the very early infall region of the PPS shows a wide range in both metallicity and quenching times. We find that sources with fast and early quenching (UDG1, UDG9, LSB5; upper left panel in Fig.~\ref{fig:SFH_alpha}) correspond to the three most metal rich sources in the mass-metallicity relation, as they appear at $>1\sigma$ of the \citep{Kirby2013} line for dwarf galaxies. The three other galaxies (UDG11, UDG12 and LSB8; lower left panel in Fig.~\ref{fig:SFH_alpha}) seem to be consistent with the standard relationship instead.

The metal-rich galaxies present a SF peak at an early epoch $t_{50} = 13.6^{+0.1}_{-0.2}$\,Gyr, with a subsequent slow and constant mass assembly phase (see Fig.~\ref{fig:SFH_alpha}, upper left panel), 
with similar $t_{90}\sim1$\,Gyr. This behaviour is atypical in galaxies located in the innermost regions of the cluster. Two of these galaxies (UDG1 and LSB5) also have low values of chemical abundance ([Mg/Fe] $\sim $ -1\,dex). These findings suggest that these galaxies might be interlopers in the PPS or have suffered inefficient environmental quenching. The remaining three galaxies in the early-infall region (UDG11, UDG12 and LSB8) show a smooth trend of the SFH (see Fig.~\ref{fig:SFH_alpha}, lower left panel), where $t_{90} = 7.6^{+0.2}_{-0.1}$\,Gyr and $t_{50} = 8.0^{+3.1}_{-0.2}$\,Gyr. It is worth noting that the SFH for these three objects in the early infall region of the cluster is quite similar to that observed for the late infallers. In particular, both groups build up their mass at a similar speed during the first phase of their mass assembly, as their $t_{50}$ estimates are consistent within uncertainties, and also, during the second phase of their formation history, driven by $\Delta t_{90}$, which are also comparable 
($\Delta t_{90,{\rm\, early}} \sim 4$\,Gyr and $\Delta t_{90,{\rm\, late}} \sim 3$\,Gyr).

\begin{figure*}
\centering
    \includegraphics[scale=0.92]{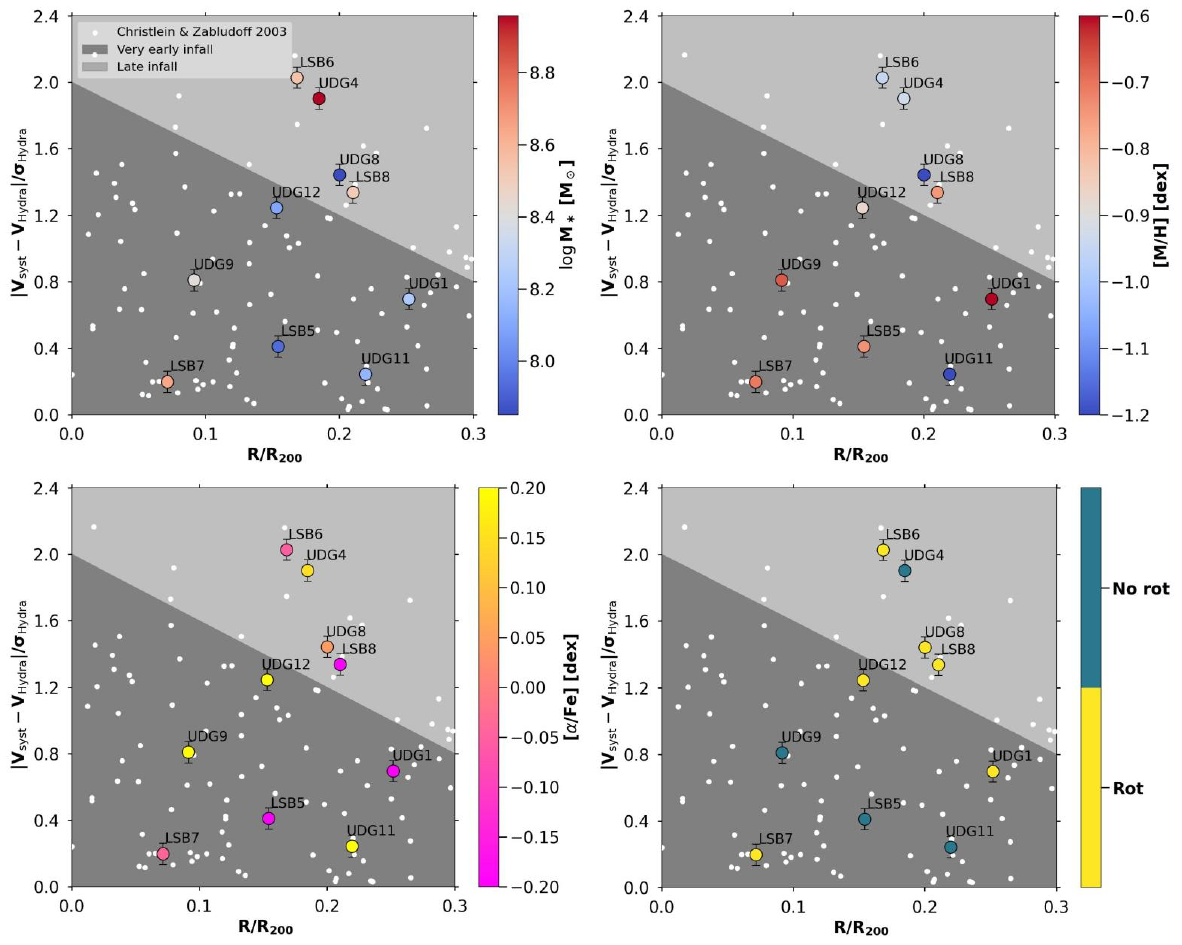}
    \caption{Global properties of LEWIS galaxies as a function of the cluster environment. The projected phase-space diagram (PPS)  shows the relative velocity versus projected radius for the LEWIS sample (coloured points), and galaxies from \cite{Christlein2003} catalogue (white circles). LEWIS galaxies are colour-coded according to the stellar mass (top-left panel), metallicity (top-right panel), chemical abundance (lower-left panel), and presence of rotation (lower-right panel). Grey-shaded regions indicate different infall stages \citep{Rhee2017, Forbes2023}: early infall (dark grey) and late infall (light grey).}
\label{fig:PPS}
\end{figure*}

\subsection{Trends within the cluster environment}
\label{in-cluster-correlations}

In Fig.~\ref{fig:PPS} we present PPS infall diagrams colour-coded with properties derived from stellar population analysis of this work and stellar kinematics from Paper~II. In the top panels of Fig.~\ref{fig:PPS}, we show the LEWIS galaxies colour-coded according to stellar mass $M_\ast$ (top left) and metallicity [M/H] (top right) derived from stellar populations analysis. We do not see any clear trend between $M_\ast$ and the galaxy location within the cluster, but we find a mild distinction in metallicity between the two groups. Except for UDG11, the majority of very early infallers show higher values of metallicities with respect to late infall galaxies. We compute the mean metallicities values in the two groups, and we find $\langle{\rm [M/H]}\rangle_{\rm early} = -0.8~\pm~0.1$\,dex for very early-infallers, and $\langle{\rm [M/H]}\rangle_{\rm late} = -1.0~\pm~0.1$\,dex for late infallers. 

In the bottom panels of Fig.~\ref{fig:PPS}, we show the LEWIS galaxies colour-coded according to the $\alpha$ abundance (bottom left) and the presence of stellar rotation (bottom right). As already pointed out in Section~\ref{sec:properties}, galaxies span a wide range of [$\alpha$/Fe] values, and we do not see a clear trend with the environment. Both early and late infallers show solar-like $\alpha$ abundances. In terms of stellar rotation, except for UDG4, galaxies in the late infall region have hints of stellar rotation, whereas in early infall regions, we find both galaxies with or without stellar rotation.


%
\subsection{Correlation with stellar kinematic properties}

From the stellar kinematic study of LEWIS galaxies (Paper~II), we found two different kinematical classes of UDG and LSB galaxies in the Hydra~I cluster. Almost half of the galaxies in the sample show significant stellar rotation. The remaining half of the galaxies do not show signs of rotation. According to the PPS, and considering that the present sample consists of only 10 objects in total, we cannot draw a firm conclusion on a possible trend of the two kinematic classes with the cluster environment. We note that the early infall region of the cluster, where most of the metal-richer objects are found, is populated by a mix of rotation-supported and pressure-supported galaxies as well (Fig.~\ref{fig:PPS}, bottom-right). Three out of the four galaxies located in the late infall region are rotation-supported. 

\begin{figure}
\centering
\includegraphics[scale=0.41]{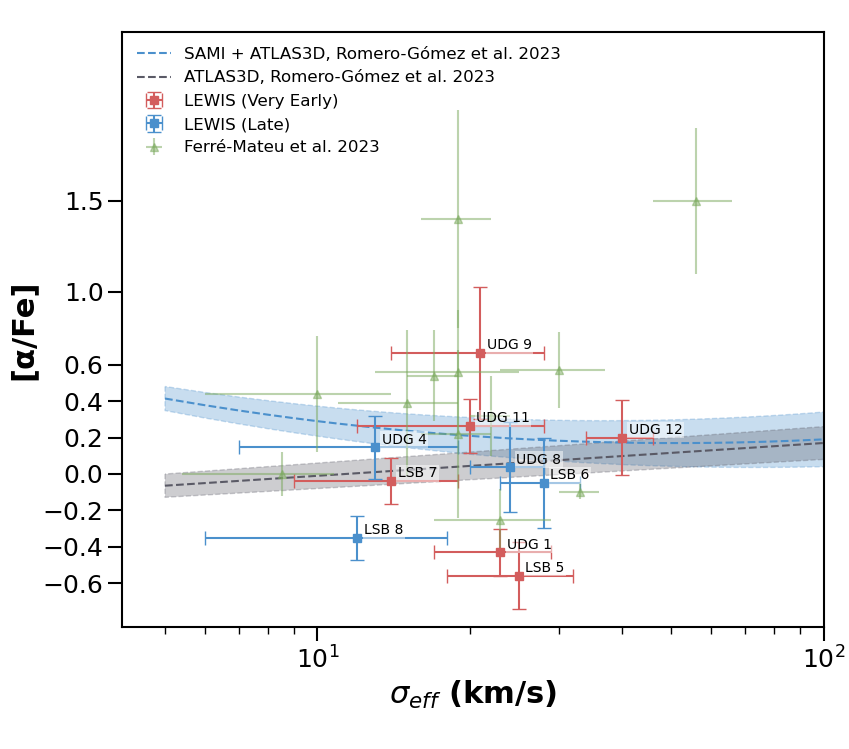}
\caption{Velocity dispersion measurements as a function of alpha abundances.   {LEWIS galaxies are colour-coded as in Fig.~\ref{fig:ages_mets_alpha}. Green triangles are UDGs analysed by \cite{FerreMateu2023}. Values of $\sigma_{\rm eff}$ for LEWIS are from Paper~II. The black dashed line and the corresponding shaded region represent the average trend of massive early-type galaxies from ATLAS$^{\rm 3D}$ survey \citep{Cappellari2011}. The light blue line and the corresponding shaded region represent the average trend ATLAS$^{\rm 3D}$ survey together with the sample of dwarf galaxies in the Fornax cluster from the SAMI survey \citep{Romero-Gomez2023}.}}
\label{fig.alpha_sigma}
\end{figure}

Fig.~\ref{fig.alpha_sigma} shows the stellar velocity dispersion values extracted from the stacked spectrum in an aperture equal to $1R_e$ ($\sigma_{\rm e}$, see Paper~II) as a function of their chemical abundance $[\alpha$/Fe].  On average, LEWIS galaxies cover the same region populated by other UDGs from the literature, expanding the range of measured [$\alpha$/Fe] from sub-solar to slightly super-solar regime. Even considering the large scatter, values for the LEWIS sample are consistent with the observed values for dwarf galaxies \citep{Romero-Gomez2023}, whose trend is marked in the figure with dashed and shaded blue region.

\begin{figure*}[t]
\centering
\includegraphics[scale=0.47]{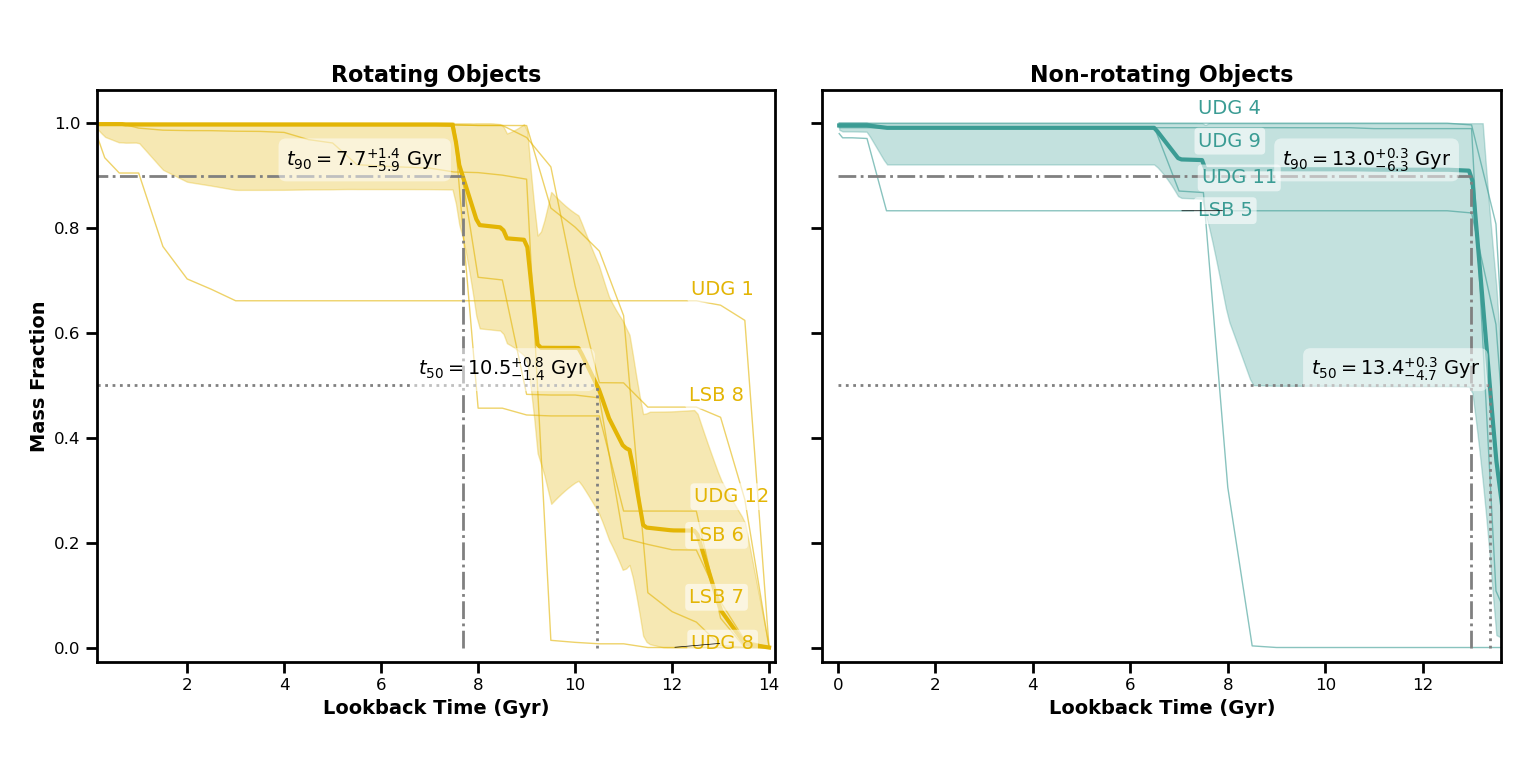}
\caption{Star-formation histories of LEWIS galaxies and stellar kinematics properties. Galaxies are separated according to the kinematic type as described in Paper~II: presence of rotation (left panel) vs. non-rotating objects (right panel). Thick, solid lines correspond to the average SFH for each panel, and shaded lines correspond to the standard deviation of the distribution. Similarly to Fig.~\ref{fig:SFH_alpha}, grey lines correspond to $t_{50}$ and $t_{90}$, respectively.}
\label{fig:kinematic_prop}
\end{figure*}

In Fig.~\ref{fig:kinematic_prop}, we show the star formation histories of LEWIS galaxies for the two kinematic types. We found that galaxies without signs of rotation present an initial fast SF phase, forming 90\% of their stellar masses in $\Delta t_{90}\sim1$\,Gyr. In contrast, galaxies that rotate show more time-extended SF, forming half of their stellar masses in $\Delta t_{50}\sim3$\,Gyr. Some individual cases have exceptional behaviour. Among the rotation-supported galaxies, UDG1 has a unique SFH, which is particularly prolonged. Interestingly, the two pressure-supported galaxies, UDG9 and LSB5, are the outliers in the mass-metallicity relation, having higher metallicities than all the other galaxies of the sample. UDG4 is the only object with a different SFH in the late infall region.

\section{Discussion: on the formation history of UDGs in the Hydra\,I cluster}
\label{sec:discussion}

The main goal of this section is to investigate the history of UDG formation in the Hydra~I cluster by comparing the observed properties of this class of galaxies, derived from the LEWIS project, 
with predictions from simulations on UDG formation. To this aim, we briefly summarise the main structural properties of the galaxies in the LEWIS sample.

By combining the stellar population properties (this work) with the stellar kinematics (Paper~II), we found that the overall structure of UDGs and LSB galaxies in Hydra~I is a function of the environment where they reside. In particular, according to the PPS (see Fig.~\ref{fig:PPS}), we found that:

\begin{itemize}
    \item Early infallers (N=6) comprise at least two different types of galaxies. We found a group of galaxies with metallicities consistent within errorbars with the mass-metallicity relation (UDG11, UDG12, LSB7). 
    Another group of galaxies, instead, has higher metallicity (UDG1, UDG9, LSB5) and is located outside the $1\sigma$ confidence region of the mass-metallicity relation. 
    The two groups show different trends of SFH. Metal-rich galaxies show a fast mass assembly phase at an early epoch, 
    with $t_{50} \sim 13.6$~Gyr, and minor SF afterwards.
    In both groups, galaxies are either pressure-supported systems (UDG9, UDG11, and LSB5) or show hints of rotation (UDG1, UDG12, and LSB7).
 
    \item Late infallers (N=4) generally present more homogeneous properties in terms of metallicity, since they seem to follow the standard mass-metallicity relationship from classical dwarfs. Among them, only UDG4 has super-solar $\alpha$-abundance and does not have hints of rotation. Almost all the other late infallers have solar-like [$\alpha$/Fe] and are rotation-supported systems. 

    \item Except for metal-rich galaxies, both early and late infallers are consistent with the mass-metallicity relation and show a similar SFH (see Fig.~\ref{fig:SFH_alpha}). 
    They build up their mass at a similar speed during the entire mass assembly period, 
    where $t_{90} \sim 8$~Gyrs for the early infallers and $t_{90}\sim 9$~Gyrs for the late infallers.
     
\end{itemize}

A summary of the properties of the UDGs/LSB populations found in Hydra~I, described above, is reported in Table~\ref{tab:summary_table}.

\subsection{How do the structural properties of the LEWIS galaxies compare with dwarfs?}

Since UDGs and LSB galaxies represent the faint end of the luminosity function of dwarf galaxies, we examined how the observed properties of the two classes of objects compare. This analysis has fundamental implications on the possible formation channels for UDGs in the Hydra~I cluster.

Firstly, according to \citet{LaMarca2022b}, colours and structural parameters of the light distribution of all UDGs and LSB galaxies are similar to those of the parent dwarf galaxy population. With the LEWIS data we can compare the stellar kinematics and populations with same quantities derived for dwarf galaxies in other clusters of galaxies\footnote{Spectroscopic data are not available for the sample of dwarf galaxies in Hydra~I cluster.}.

According to the standard picture of dwarf galaxy evolution \citep{Mistani16}, very early-infallers are expected to have an increased $\alpha$-abundance pattern, low metallicity, old ages, and a large amount of GCs \citep{Forbes2023}. In previous literature, UDG and LSB metallicities have been shown to have only a weak correlation with clustercentric distance (e.g, \citealp{Romero-Gomez2023}, \citealp{Ferre-Mateu2018}), yet low number statistics and the very limited number of studied environments (mostly Coma, Virgo and Perseus, see \citealp{FerreMateu2023, Gannon2024} and references therein) have hindered strong conclusions.

Overall, in the Hydra~I cluster, we observe that late infallers exhibit properties similar to those of regular dwarfs. Conversely, early infallers display a larger scatter in all properties (metallicity, $\alpha$-enrichment, and stellar kinematics), indicating that they likely contain galaxies that have undergone at least two distinct evolutionary paths. In particular, based on the star-formation history shape, or more specifically, the [$\alpha$/Fe] ratio, two major populations are found.

A sub-population of the LEWIS sample resembles the properties of regular dwarfs (LSB5, LSB6, LSB7, LSB8, UDG1, UDG8) with their low-to-`around-solar' [$\alpha$/Fe] ratios and extended SFHs. These could include the LSB tail of either high-metallicity dwarf-spheroidals (UDG1, LSB6, LSB7, LSB8, possibly slightly enriched by their medium yet supported by their rotation) or SF-inefficient dwarf-ellipticals (LSB5). These are consistent with the mass-metallicity relationship for dwarfs, perhaps slightly enriched by the Hydra-I environment. According to the findings of Paper~II, most of these galaxies, regardless of their strict UDG/LSB classification, appear to be DM-rich and exhibit rotation. In particular, as shown in Fig.5 of Paper~II, the projected specific angular momentum ($\lambda_{\rm R}$) found for these objects are comparable with typical values derived for dwarf galaxies of similar stellar masses. The rest of the LEWIS sample is consistent with the recent UDG literature, showing solar-like or super-solar [$\alpha$/Fe] patterns and fast mass assembly.

\subsection{Comparing the observed properties with predictions from simulations}

The LEWIS sample consists of galaxies located within 0.4$R_{\rm vir}$ of Hydra~I, meaning we are mapping a high-density environment. Therefore, we focus solely on comparison with theoretical predictions of UDGs in clusters.  

Using the Illustris-TNG simulations, \citet{Sales2020} and \citet{Benavides2023} provided detailed properties on both stellar kinematics and stellar populations in cluster UDGs, under the assumption that all their UDGs are dispersion-supported systems. In this scenario, the T-UDGs, i.e., galaxies transformed into UDGs through tidal stripping, are predicted to have higher metallicity and lower velocity dispersion than dwarf galaxies of comparable stellar mass and the B-UDGs, which are field galaxies born as UDGs that later join the cluster. Since almost half of the UDGs in the Hydra~I cluster are rotation-supported, a direct comparison with predictions from \cite{Sales2020} can be performed only for the dispersion-supported galaxies, considering those targets having a constrained fit for the stellar kinematics (see Sec.~\ref{sec:data_obs} and Paper~II). These are UDG4, UDG9, UDG11, LSB5 (see Table~\ref{tab:stellar_pop}). Examining the mass-metallicity relation in Fig.~\ref{fig:MZR}, UDG9 and LSB5 are outliers, exhibiting higher metallicity than those for dwarf galaxies of similar stellar masses. None of these two galaxies shows a velocity dispersion $\sigma_e$ lower than the typical values for dwarf galaxies, as predicted by the Faber-Jackson relation (see Fig.4 in Paper II). However, given the large scatter of the simulated UDGs in this relation, the tidal-stripping origin for these galaxies cannot be excluded. Both LSB5 and UDG9 are located in the early-infall region of the PPS, where tidal interactions between galaxies are clearly detected \citep{Spavone2024}, and where T-UDGs are expected to be found.

UDG4 and UDG11 have $\sigma_{\rm eff}$ and metallicity that are consistent with the expected Faber-Jackson and mass-metallicity relations. Combining velocity dispersion measurements with metallicity estimates, these LEWIS galaxies might resemble the B-UDGs in the framework proposed by \citet{Sales2020}. However, the number density of B-UDGs in these simulations is predicted to be larger in the cluster outskirts, with a half-number radius of about 0.5 of the virial radius. Since the LEWIS sample covers $\sim 0.4$~$R_{\rm vir}$ of the Hydra~I cluster, not many B-UDGs should be expected in our sample.

UDG1, instead, has hints of stellar rotation, and thus cannot be directly compared with \cite{Sales2020} prediction. This galaxy is a peculiar case. It is located in the very early infall region of the PPS, but differently from almost all other UDGs and LSBs in this region, it has a slow and constant mass assembly phase with a second, more recent starburst at $\sim 2$ Gyr. This is hard to reconcile with a standard tidal origin or a failed formation channel.

Results from UDGs in the TNG50 simulations provide constraints on metallicity \citep{Benavides2023, Benavides2024}, which can be compared with estimates derived in this work. Authors argued that these UDGs, embedded in a dwarf-like DM halo, might originate from a dwarf progenitor whose stellar content is puffed up due to a high-spin DM halo. Simulated quenched UDGs are on average metal-poor, ranging from [Fe/H]$
\sim$-1.3 [dex] at $M_\ast\sim10^{7.5}M_\odot$ and [Fe/H]$\sim$-0.9 [dex] at $M_\ast\sim10^{9}M_\odot$, and are consistent with the MZR relation for dwarf galaxies \cite{Kirby2013}, even considering the large scatter of values.

The metallicity of the LEWIS galaxies spans a range $-1.3 \leq \langle{\rm [M/H]}\rangle \leq -0.5$\,dex, with stellar masses $M_\ast \leq 10^{9}M_\odot$ (see Fig.~\ref{fig:MZR}). Taking into account the error estimates, almost all LEWIS galaxies, except the three objects with the highest metallicities, show comparable values to those predicted by the simulations cited above. Therefore, considering the small scatter around the mass-metallicity relation, and the similarities in other structural properties (integrated colours, light distribution, SFH, $\alpha$-enhancement, stellar kinematics), 
all of them might reconcile with a puffed-up origin.

In the mass-metallicity relation, none of the galaxies in the LEWIS sample shows metallicity lower than  $\sim -1.5$\,dex (see Fig.~\ref{fig:ages_mets_alpha}). As suggested by \citet{Buzzo2024} and \citet{FerreMateu2023}, very low metallicities, consistent with the evolving mass-metallicity relationship at redshift $z\sim 2.2$, are expected for the UDGs formed as the `failed-galaxy' scenario (see Fig.~\ref{fig:MZR}). However, we cannot rule out their presence at the outskirts of the cluster, beyond the area covered by LEWIS.

\begin{table*}
\centering
\caption{Stellar populations properties of early and late UDG populations in Hydra~I.}
\renewcommand{\tabcolsep}{0.4cm}
\renewcommand{\arraystretch}{1.35}
\footnotesize
\begin{tabular}{c c c}
\hline
\hline
Properties & Very early infallers  & Late infallers \\
(1)   & (2)   & (3)\\
\hline
Number of galaxies & 6 galaxies  &  4 galaxies  \\
\hline
Metallicity  & $[{\rm M/H}]=-0.8\pm0.1$\,{\rm dex}  &  ${\rm [M/H}]=-1.0\pm0.1$\,{\rm dex}  \\
$\alpha$ abundances & $[\alpha/{\rm Fe}]= 0.0\pm0.2\,{\rm dex}$  & $[\alpha/{\rm Fe}]= -0.1\pm0.1\,{\rm dex}$ \\
Mass-Metallicity relation$^{(a)}$ & {\rm dwarf-like \& metal-richer} & {\rm dwarf-like} \\
\hline
\multirow{2}{*}{Star formation history$^{(b)}$} & {\rm dwarf-like}: $\Delta t_{50} \sim 6$\,Gyr , $\Delta t_{90} \sim 0.5$\,Gyr & {\rm dwarf-like}: $\Delta t_{50} \sim 5$\,Gyr , $\Delta t_{90} \sim 1$\,Gyr \\
 &  {\rm metal-richer}: $\Delta t_{50} \sim 0.5$\,Gyr , $\Delta t_{90} \sim 12$\,Gyr &   \\
\hline
Stellar rotation$^{(c)}$ & Rotation \& pressure-supported &  Rotation-supported \\
\hline
\end{tabular}
\label{tab:summary_table}
\tablefoot{  {Column 1 reports the properties of stellar population analysis from this work and stellar kinematics analysis from Paper II. Columns 2 and 3 report the values for the very early and late infallers. $(a)$: we refer to galaxies in LEWIS for which the value of metallicity is similar (dwarf-like) or higher (metal-richer) than metallicity of dwarf galaxies of similar stellar mass according to the \cite{Kirby2013} relation. $(b)$ we are not considering UDG4 since it has a peculiar SFH (see Fig.~\ref{fig:SFH_alpha}, top left panel). $(c)$: in the late-infallers, three out of four galaxies have hints of stellar rotation, while the remaining one is pressure-supported. The reported classification refers to the majority of the objects in this class, although the sample size is not statistically significant.}}
\end{table*}

\section{Summary and conclusive remarks}\label{sec:conclusion}

In this paper, we have presented the stellar population properties for a homogeneous sample of UDG and LSB galaxies in the Hydra~I cluster of galaxies observed with the MUSE integral-field spectrograph at ESO-VLT, as part of the LEWIS large programme \citep{Iodice2023}. We have derived the age, metallicity, and  $\alpha$-abundances for 10 galaxies in the LEWIS sample (see Table~\ref{tab:stellar_pop}).  Overall, we have found that these galaxies exhibit age and metallicity consistent with typical values for dwarf galaxies of comparable stellar mass, including literature estimates for UDGs. The stellar population properties derived in this work, together with the stellar kinematics presented in Paper~II, are analysed as a function of the cluster environment using the projected phase space (PPS).

In the late-infall region of the PPS, we identify a single population of galaxies with the lowest metallicities, nearly all of which are rotation-supported systems. In contrast, the early-infall region contains two types of galaxies: roughly half have metallicities consistent with the dwarf galaxy mass–metallicity relation, while the other half show higher metallicities (with $\langle{\rm [M/H]}\rangle \geq -1.0$\,dex), lying outside the $1\sigma$ confidence region of that relation. Both sub-samples include rotation- and pressure-supported systems. Galaxies in this region of the cluster also display different timescales for stellar mass assembly. Metal-rich galaxies reached 50\% of their stellar mass in less than 1~Gyrs before a shutdown in SF. UDG1 (and possibly LSB5) do show a second, localized peak at intermediate ages (1-2~Gyrs). All other galaxies exhibit a SFH similar to that found for galaxies in the late-infall region.

By comparing the observed properties for LEWIS galaxies with possible formation channels, we suggest that most of the UDGs/LSBs inside the $0.4R_{\rm vir}$ of the Hydra~I cluster are likely puffed-up dwarfs. 
We propose that the high-metallicity systems have undergone significant stellar evolution driven by tidal processes, likely through interactions with nearby bright and massive galaxies \citep[see][and references therein]{Spavone2024}. This interpretation is supported by their location in the cluster core and North group, where deep imaging has revealed clear signatures of ongoing gravitational interactions, such as tidal tails and disrupted dwarf galaxies. Finally, we do not find any galaxies consistent with the “failed-galaxy” scenario, primarily due to the absence of very low-metallicity systems in our sample. However, we cannot rule out the existence of such systems in the cluster outskirts, beyond the area covered by LEWIS.

In this work, we have attempted for the first time to correlate the spatial information from stellar kinematics (from Paper II) with age, metallicity and SFH. We expect to place even tighter constraints on the formation histories of UDGs in Hydra~I by combining the structural properties derived so far (from stellar populations and stellar kinematics) with the globular cluster (GC) content \citep{Mirabile2025}.

\begin{acknowledgements}
We wish to thank the anonymous Referee whose comments helped us to improve the clarity of the manuscript. Based on observations collected at the European Southern Observatory under ESO programmes 108.222P.001, 108.222P.002, 108.222P.003. The authors wish to thank the Instituto de Astrofísica de Canarias for their hospitality and mentorship during January to April. Additionally, we are grateful to Jonah Gannon, Maria Luisa Buzzo, Thomas Puzia, and Alexandre Vazdekis for their helpful discussions and valuable suggestions. E.I. acknowledges support by the INAF GO funding grant 2022-2023. 
EI, EMC and MP acknowledge the support by the Italian Ministry for Education University and Research (MIUR) grant PRIN 2022 2022383WFT “SUNRISE”, CUP C53D23000850006.
EMC acknowledges the support from MIUR grant PRIN 2017 20173ML3WW-001 and Padua University grants DOR 2021-2023.
GD acknowledges support by UKRI-STFC grants: ST/T003081/1 and ST/X001857/1.
A.F.M. has received support from RYC2021-031099-I and PID2021-123313NA-I00 of MICIN/AEI/10.13039/501100011033/FEDER,UE, NextGenerationEU/PRT
DF thanks the ARC for support via DP220101863 and DP200102574. 
J.F-B acknowledges support from the PID2022-140869NB-I00 grant from the Spanish Ministry of Science and Innovation. 
J.H. and E.I. acknowledge the financial support from the Visitor and Mobility program of the Finnish Centre for Astronomy with ESO (FINCA).
This work is based on the funding from the INAF through the GO large grant in 2022, to support the LEWIS data reduction and analysis (PI E. Iodice). 
The authors thank \citet{Gannon2024} for the compilation of their catalogue of UDG spectroscopic properties. The catalogue includes data from: \citet{mcconnachie2012, vanDokkum2015, Beasley2016, Martin2016, Yagi2016, MartinezDelgado2016, vanDokkum2016, vanDokkum2017, Karachentsev2017, vanDokkum2018, Toloba2018, Gu2018, Lim2018, RuizLara2018, Alabi2018, Ferre-Mateu2018, Forbes2018, Martin2019, Chilingarian2019, Fensch2019, Danieli2019, vanDokkum2019, torrealba2019, Iodice2020, Collins2020, Muller2020, Gannon2020, Lim2020, Muller2021, Forbes2021, Shen2021, Ji2021, Huang2021, Gannon2021, Gannon2022, Mihos2022, Danieli2022, Villaume2022, Webb2022, Saifollahi2022, Janssens2022, Gannon2023, FerreMateu2023, Toloba2023, Shen2023}. 
The authors acknowledge the use of the following Python scripts: 
{\sc ASTROPY} \citep{astropy:2013, astropy:2018}, {\sc MATPLOTLIB} \citep{matplotlib}, {\sc MPDAF} \citep{Bacon2016, Piqueras2017}, {\sc NUMPY} \citep{numpy}, {\sc PHOTUTILS} \citep{photutils}, {\sc SCIPY} \citep{SciPy}, and {\sc ZAP} \citep{Soto2016}.

\end{acknowledgements}

%
%

\bibliography{LEWIS_bib}{}
\bibliographystyle{aa}

\appendix

\section{Line indices}
\label{appendix-lineindices}

\begin{figure*}[!h]
\centering
\includegraphics[scale=0.4]{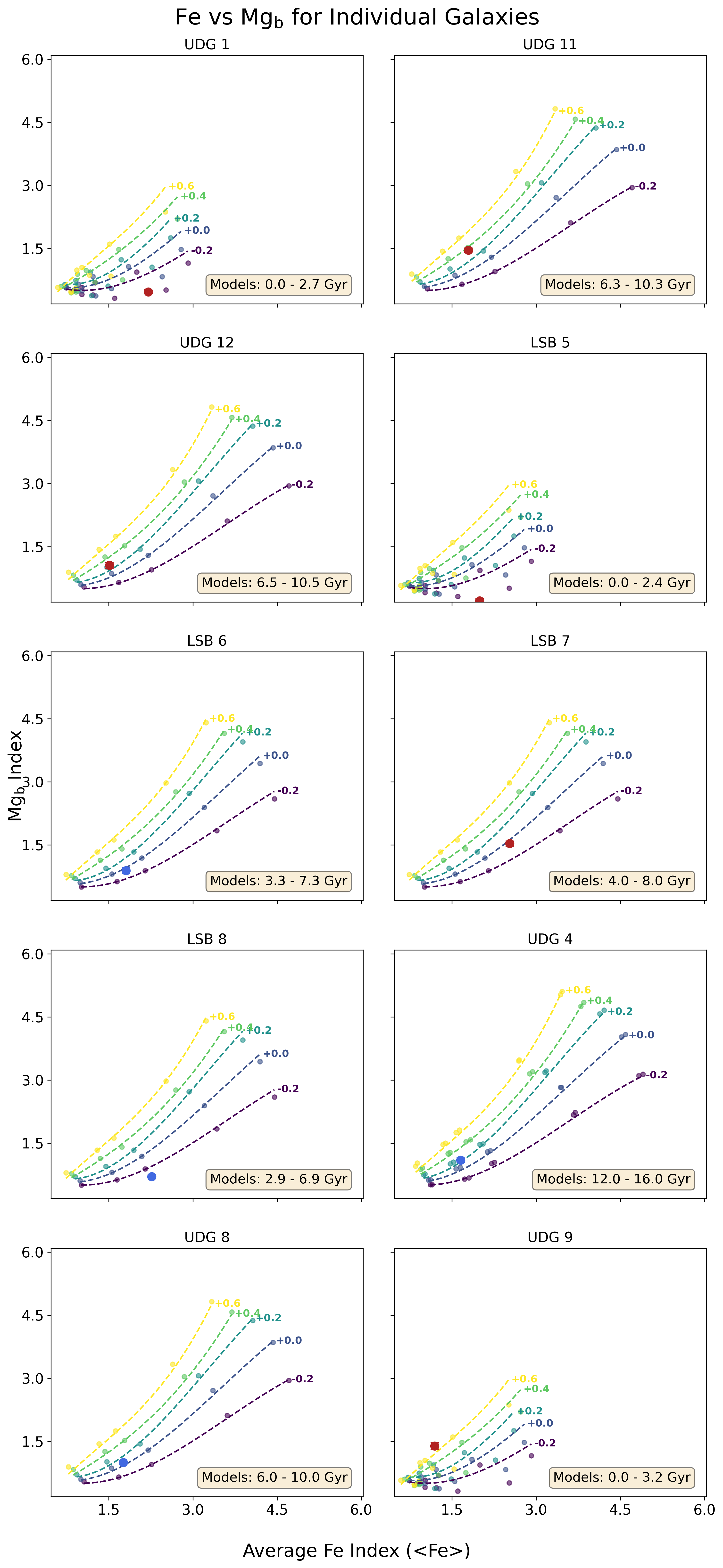}
\caption{  {Line indices $\langle {\rm Fe} \rangle$ and $\langle {\rm Mg_b} \rangle$ for sMILES models and LEWIS galaxies. In each panel, the label shows the subset of sMILES models (coloured lines and circles) restricted in age and metallicity used to compute line indices and infer the values of [Mg/Fe] for the LEWIS galaxies, colour-coded as in Fig.~\ref{fig:ages_mets_alpha}.}}
\label{fig.Licks-MgFe}
\end{figure*}

In this section, we show the model grids used for the computation of the [Mg/Fe] pattern and address the potential caveats of using the standard E-MILES models. 

Fig.~\ref{fig.Licks-MgFe} shows a selection of sMILES model grids for all galaxies in our high-quality sample: for each one, we restrict the age and metallicity of each grid to a region centred at the {\tt ppxf} solution of the galaxy for both their age and metallicity. Comparing the datapoints to an interpolated (or extrapolated) trend for the grid, we compute an estimate of [Mg/Fe].

\begin{figure*}[!h]
\centering
\includegraphics[scale=0.42]{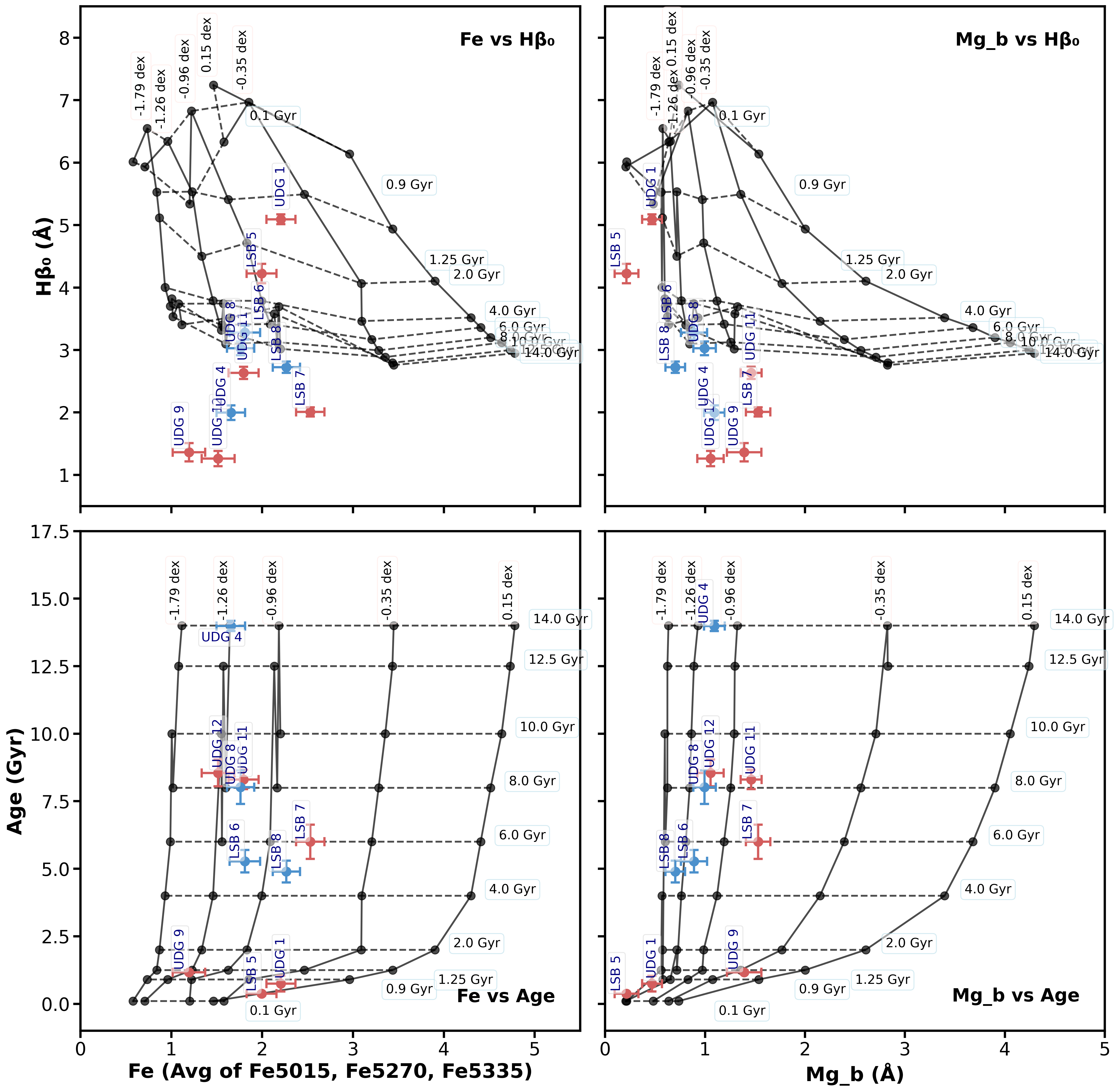} 
\caption{Top row: E-MILES base models compared to line indices datapoints from our sample. Bottom row: E-MILES model grids used to compute the [Mg/Fe] enhancement, using the light-weighted age of the models and {\tt ppxf} to make the grids orthogonal, and this is better defined in the old and metal-poor regime. LEWIS galaxies are colour-coded as in Fig.~\ref{fig:ages_mets_alpha}}
\label{fig.Licks-age}
\end{figure*}

The top panel of Fig.~\ref{fig.Licks-age} shows our data points on top of the E-MILES models. The bottom panel shows the actual model grids (E-MILES+{\tt ppxf}) to derive the [Mg/Fe] pattern, using light-weighted ages as a more robust age indicator. We note that E-MILES templates are not being used for any {\tt ppxf} fits and are solely to have another, partly independent measurement of the alpha enhancement. We strictly use variable [$\alpha$/Fe] sMILES templates, as these allow us to cover a larger space in the H$\beta_0$ distribution and therefore to minimise any effect of template mismatch. We caution that two sources have an extremely low H$\beta_0$ intensity and are not perfectly covered by any template (UDG9 and UDG12). However, the strength of the metallic lines allows us to still constrain the metallicity and alpha content of these galaxies via the alpha-enhanced sMILES templates. The age, though in a difficult position, can be constrained by the use of {\tt ppxf}. We interpret the positioning of these sources within the grids as the effect of high levels of [$\alpha$/Fe], low S/N and a possible, minor emission component, hard to decouple given the quality of these spectra.

Overall, we find the results for the [Mg/Fe] computation are fully consistent between the three methods ({\tt ppxf}-derived, E-MILES derived, and sMILES-derived). Three minor exceptions are found: UDG1, LSB5 and   {LSB8} have an extremely low alpha-enhancement, going slightly beyond the template limits at the low end with the sMILES method (Fig.~\ref{fig.Licks-MgFe}) and are at the corners of the grid following the method of \cite{Vazdekis2015} and \cite{FerreMateu2023} using the E-MILES templates (Fig.~\ref{fig.Licks-age}). Thus, we are forced to slightly extrapolate outside the model grids for their computation, as the sMILES models (and thus, their {\tt ppxf} results) have a hard limit at a sub-solar enhancement of $-0.2$ dex.

\section{Stellar population properties and star-formation histories}
\label{appendix}

\begin{figure*}[h]
\centering
\includegraphics[scale=0.44, clip]{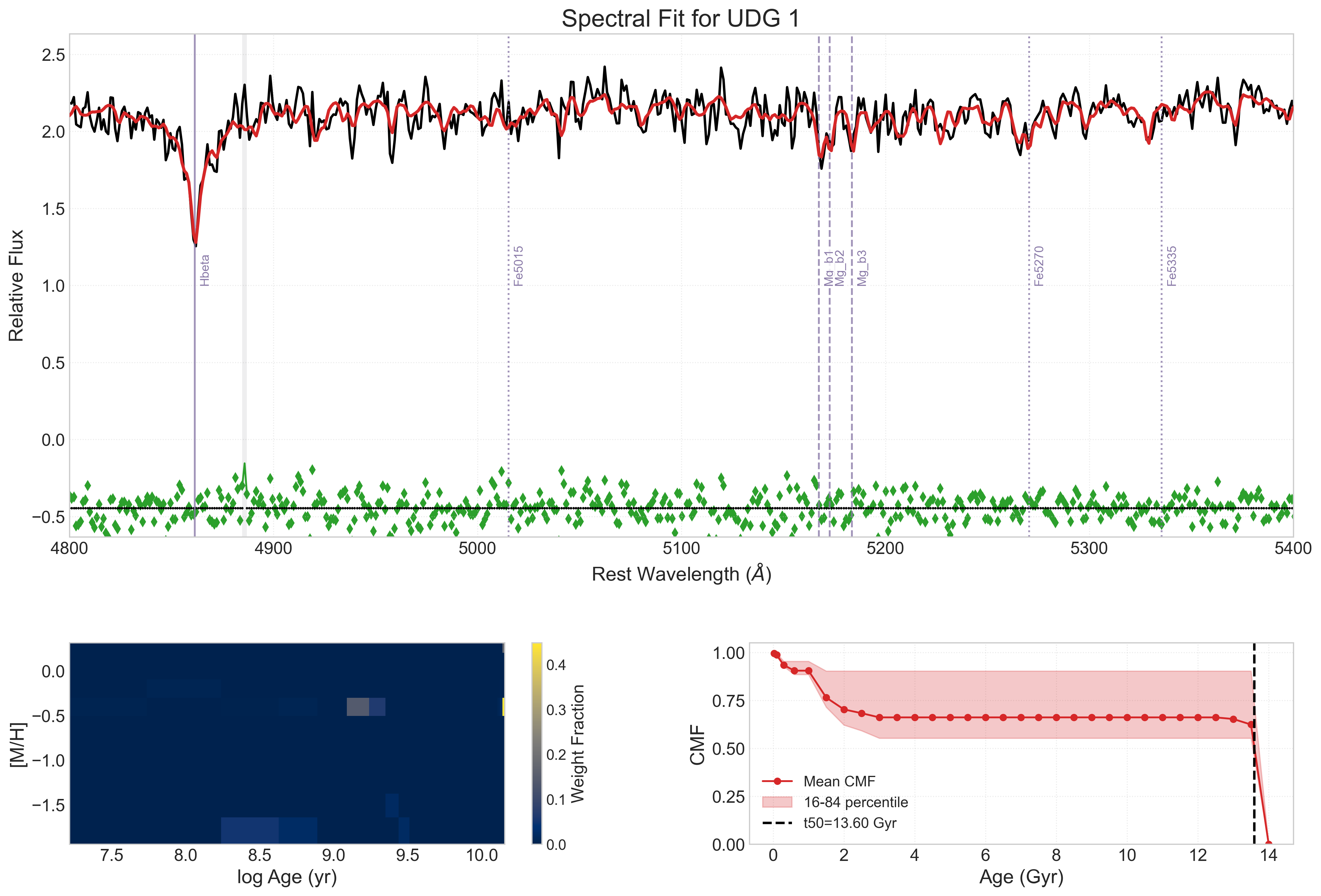}
\caption{Stellar population properties and star-formation history of UDG1. Top panel: MUSE 1\,$R_{\rm e}$ stacked spectrum (black solid line) for UDG1. The main absorption features are marked in purple. The red solid line represents the bestfit obtained with {\tt ppxf}. Green points are the residuals between the observed spectrum and its best-fit template spectrum. The grey areas are the masked regions excluded from the fit. Bottom left panel: Age-metallicity distribution of the MC iterations of UDG1. Bottom right panel: Cumulative mass fraction, based on the MC iterations. $t_{50}$ marks the time at which the 50\% of the galaxy mass was formed.}
\label{fig.UDG1_panels}
\end{figure*}

\begin{figure*}[t]
\centering
\includegraphics[scale=0.44, clip]{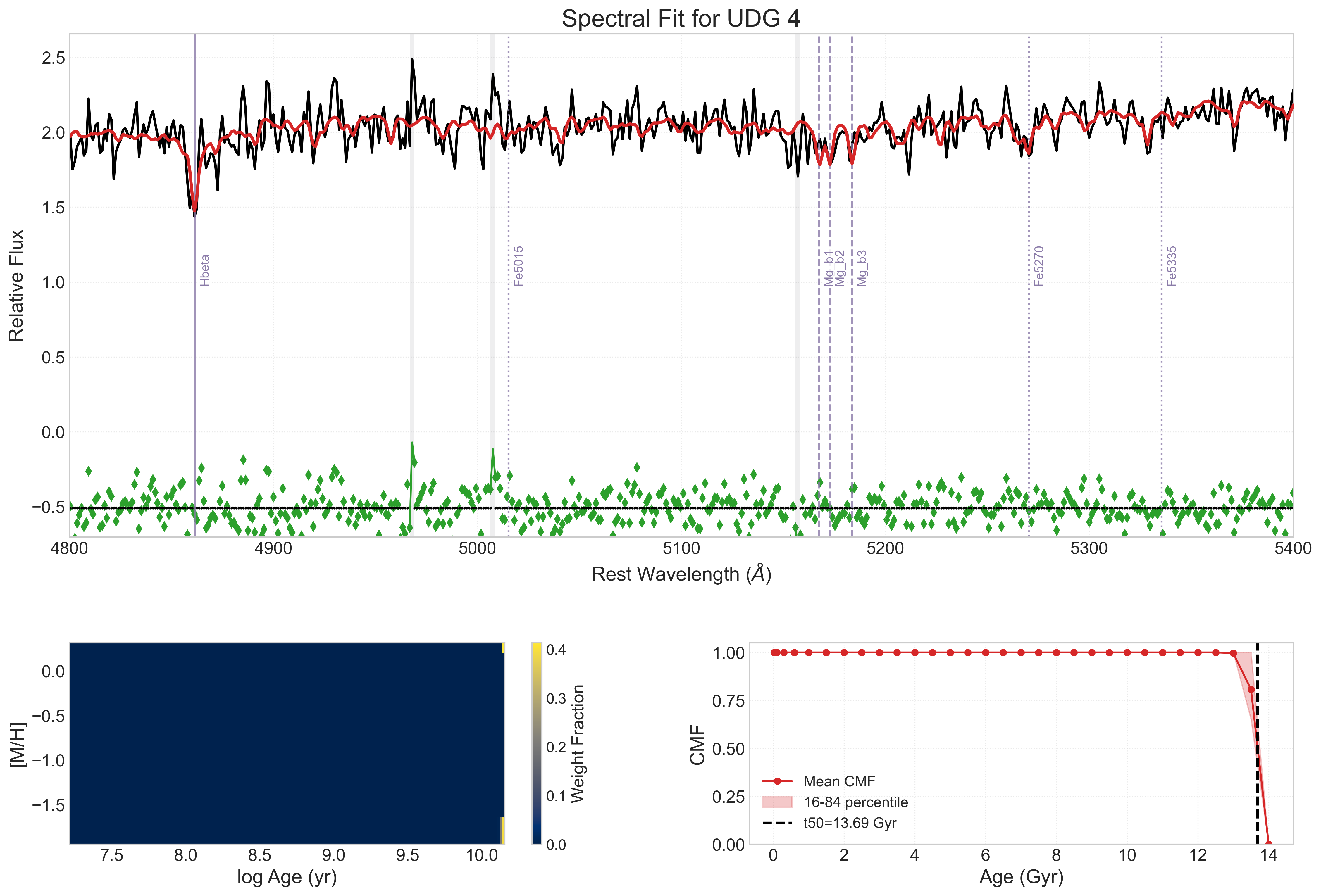}
\caption{Same as Fig.~\ref{fig.UDG1_panels}, but for UDG4.}
\label{fig.UDG4_panels}
\end{figure*}

\begin{figure*}[t]
\centering
\includegraphics[scale=0.44, clip]{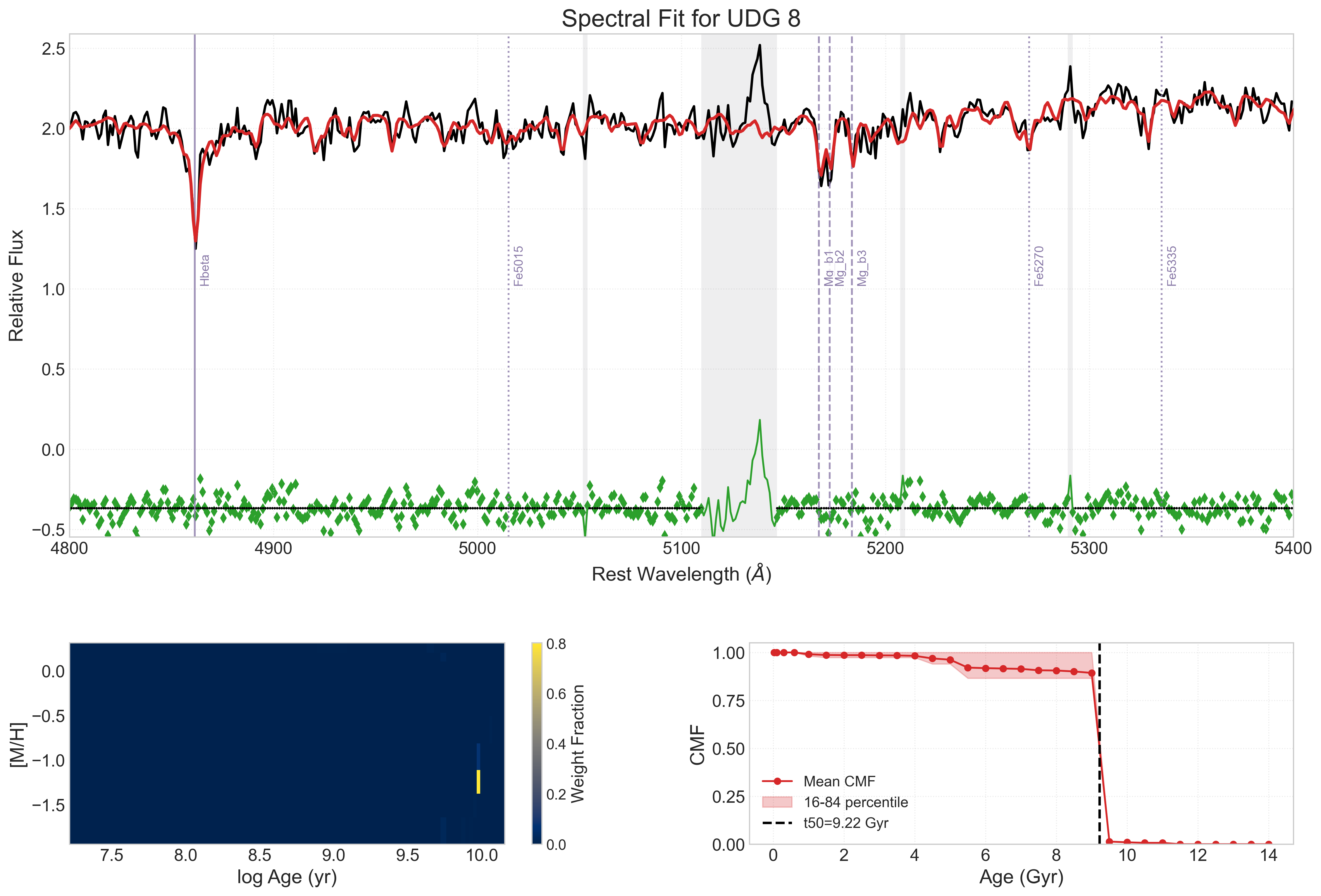}
\caption{Same as Fig.~\ref{fig.UDG1_panels}, but for UDG8.}
\label{fig.UDG8_panels}
\end{figure*}

\begin{figure*}[t]
\centering
\includegraphics[scale=0.44, clip]{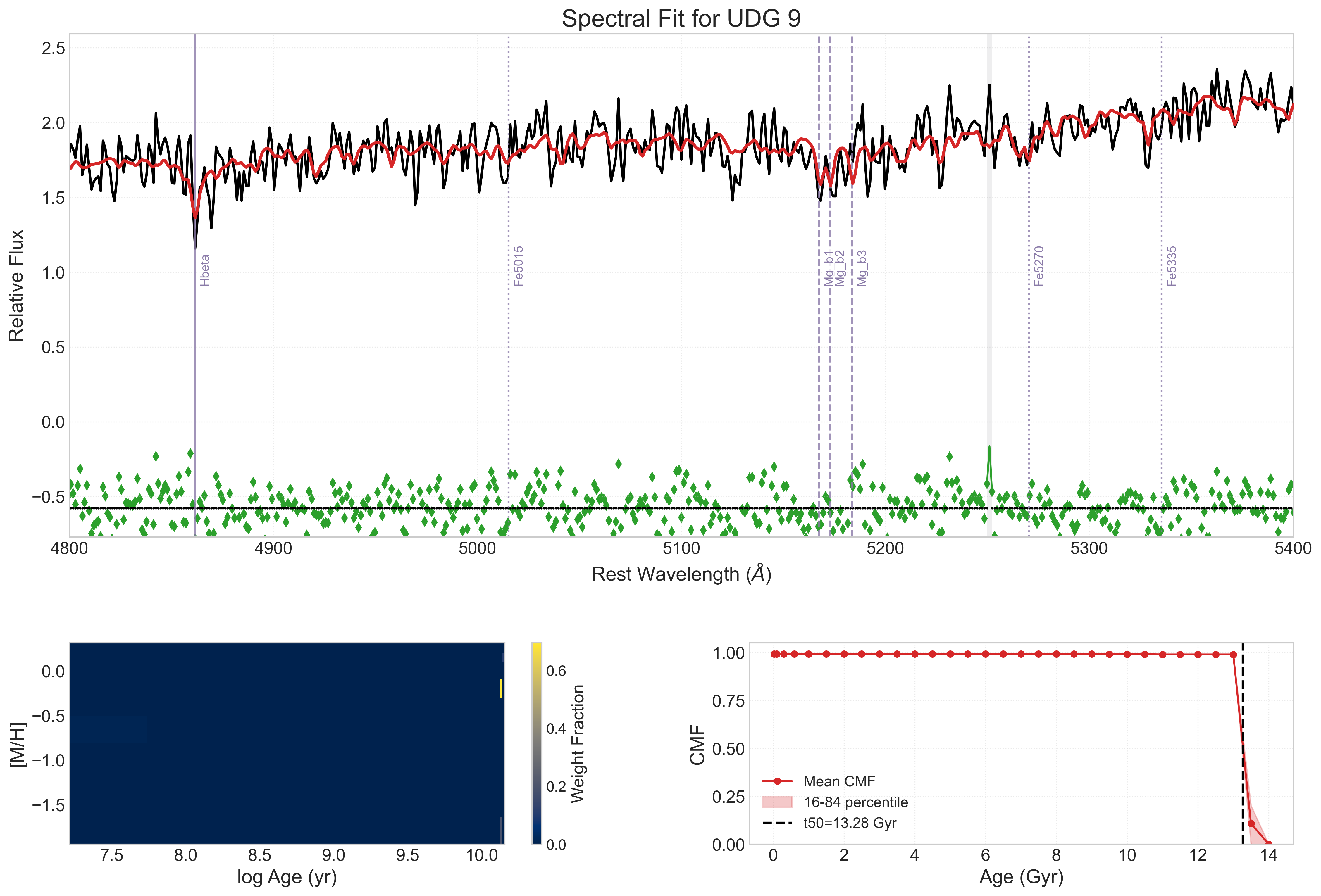}
\caption{Same as Fig.~\ref{fig.UDG1_panels}, but for UDG9.}
\label{fig.UDG9_panels}
\end{figure*}

\begin{figure*}[t]
\centering
\includegraphics[scale=0.44, clip]{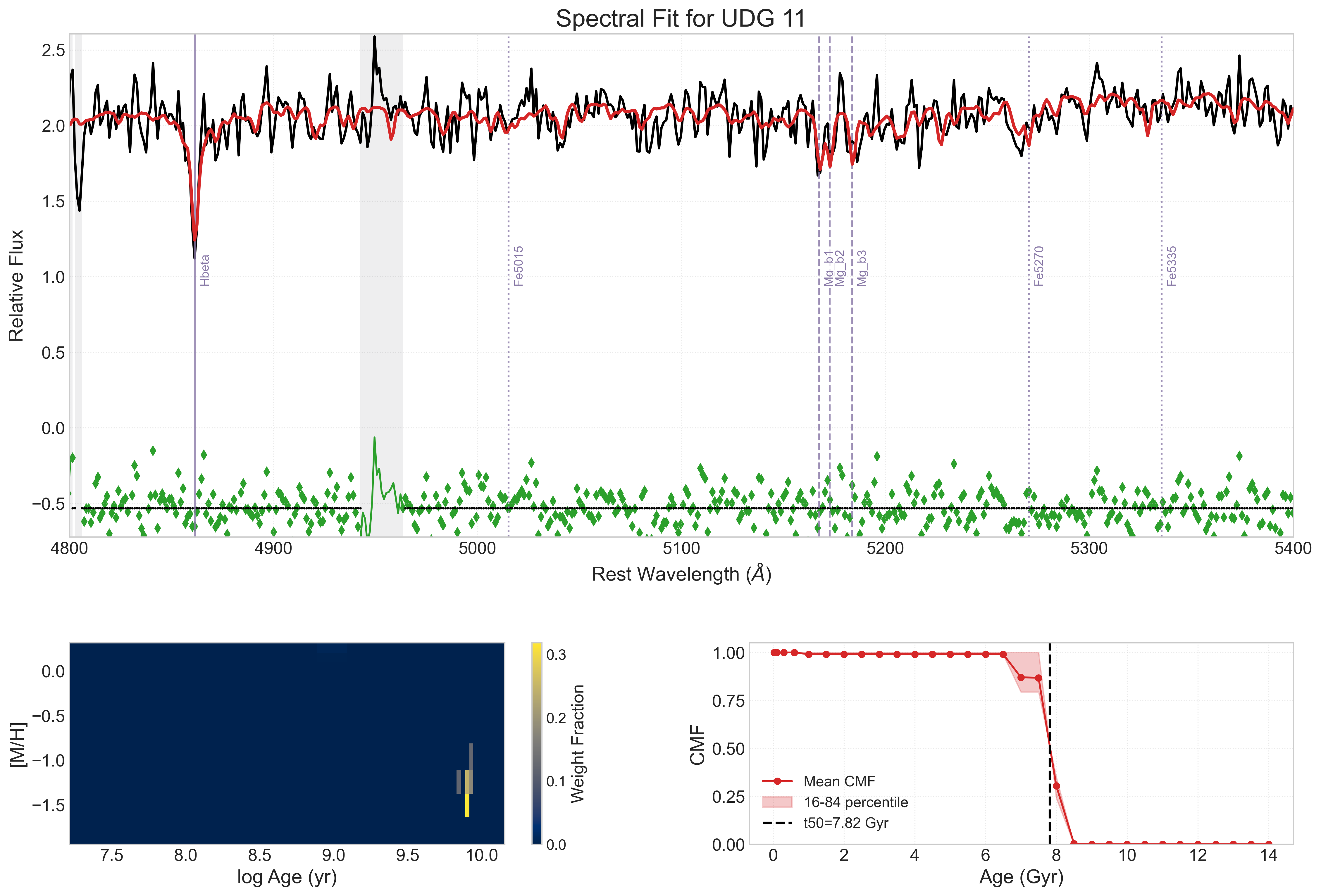}
\caption{Same as Fig.~\ref{fig.UDG1_panels}, but for UDG11.}
\label{fig.UDG11_panels}
\end{figure*}

\begin{figure*}[t]
\centering
\includegraphics[scale=0.44, clip]{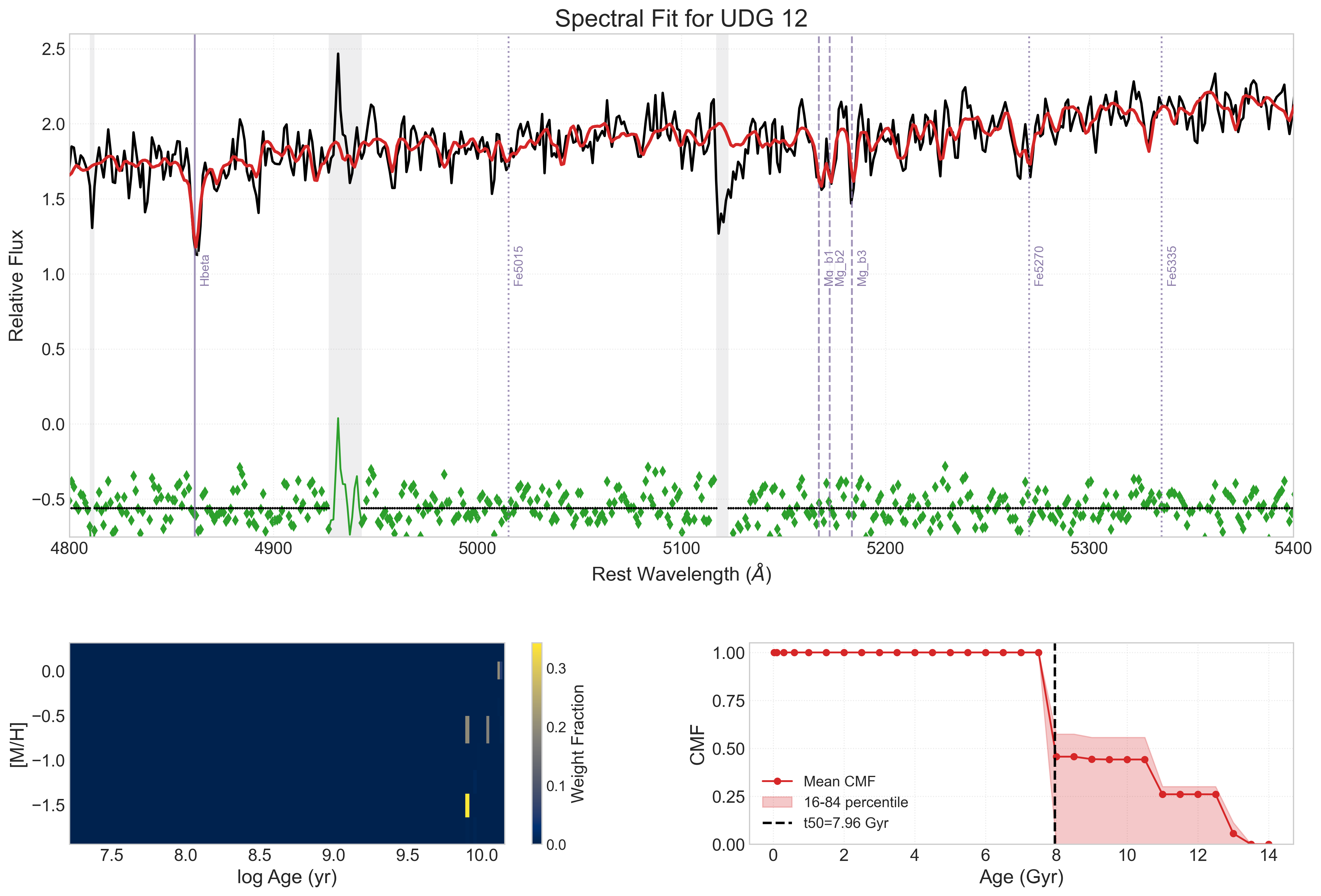}
\caption{Same as Fig.~\ref{fig.UDG1_panels}, but for UDG12.}
\label{fig.UDG12_panels}
\end{figure*}

\begin{figure*}[t]
\centering
\includegraphics[scale=0.44, clip]{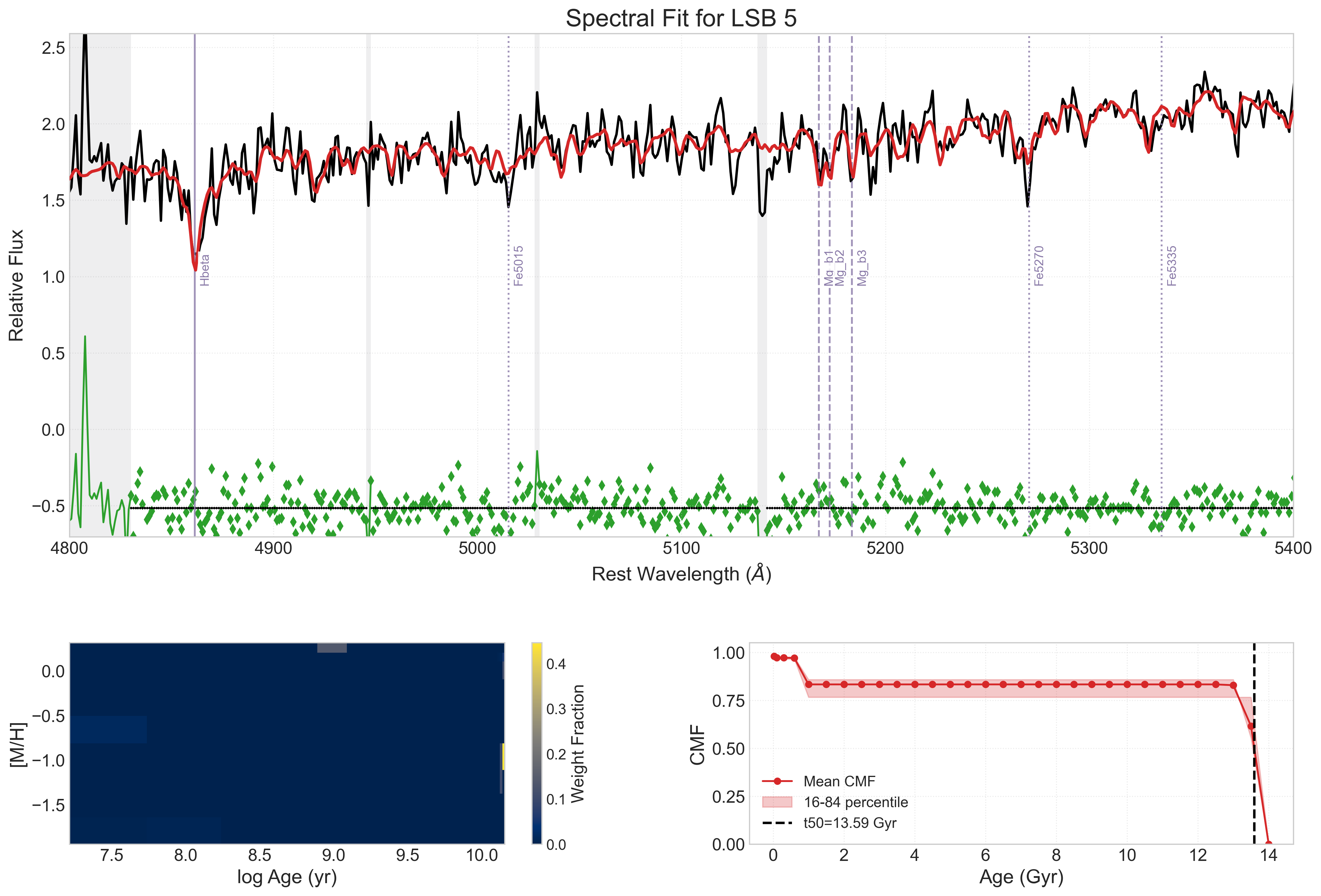}
\caption{Same as Fig.~\ref{fig.UDG1_panels}, but for LSB5.}
\label{fig.LSB5_panels}
\end{figure*}

\begin{figure*}[t]
\centering
\includegraphics[scale=0.44, clip]{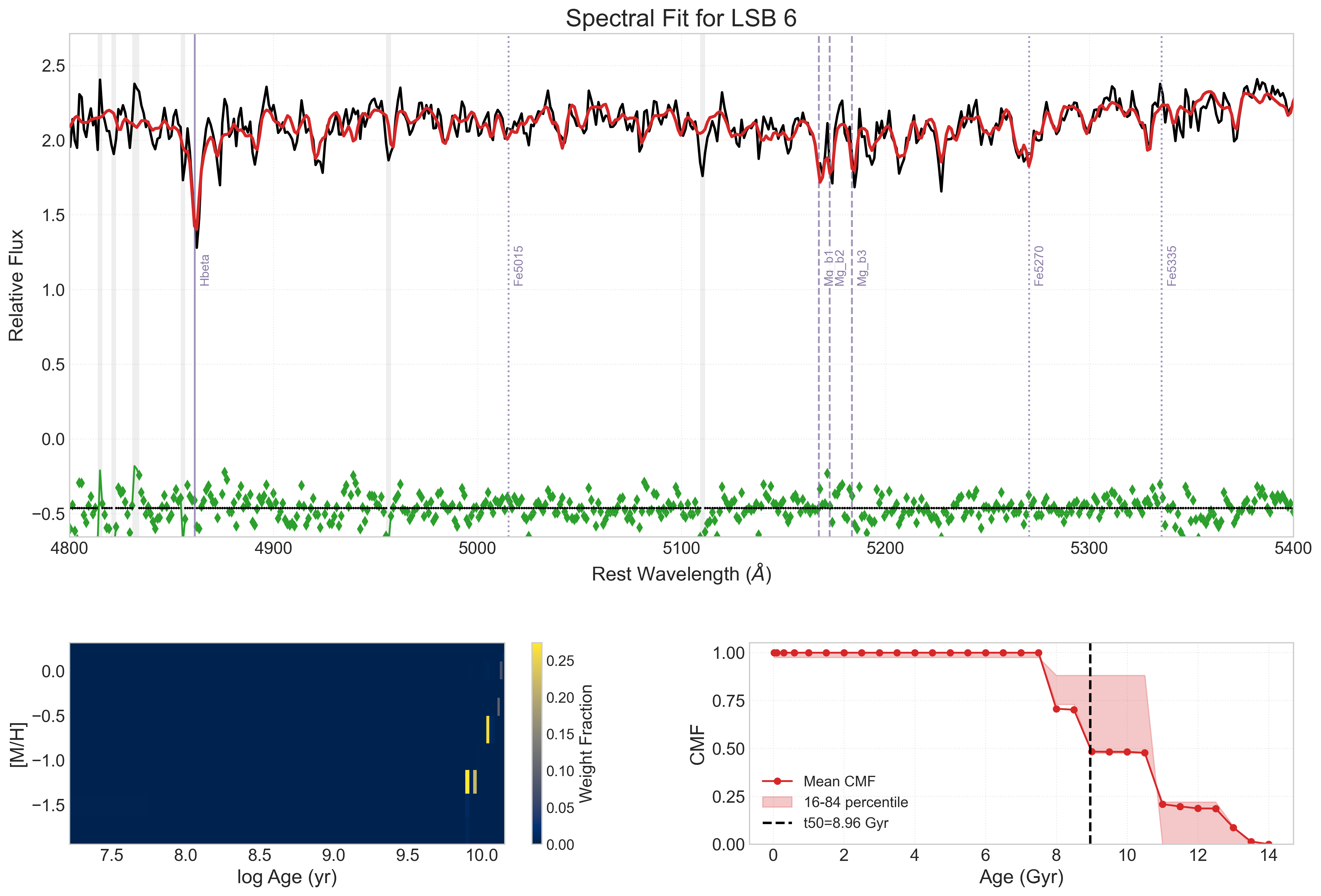}
\caption{Same as Fig.~\ref{fig.UDG1_panels}, but for LSB6.}
\label{fig.LSB6_panels}
\end{figure*}

\begin{figure*}[t]
\centering
\includegraphics[scale=0.44, clip]{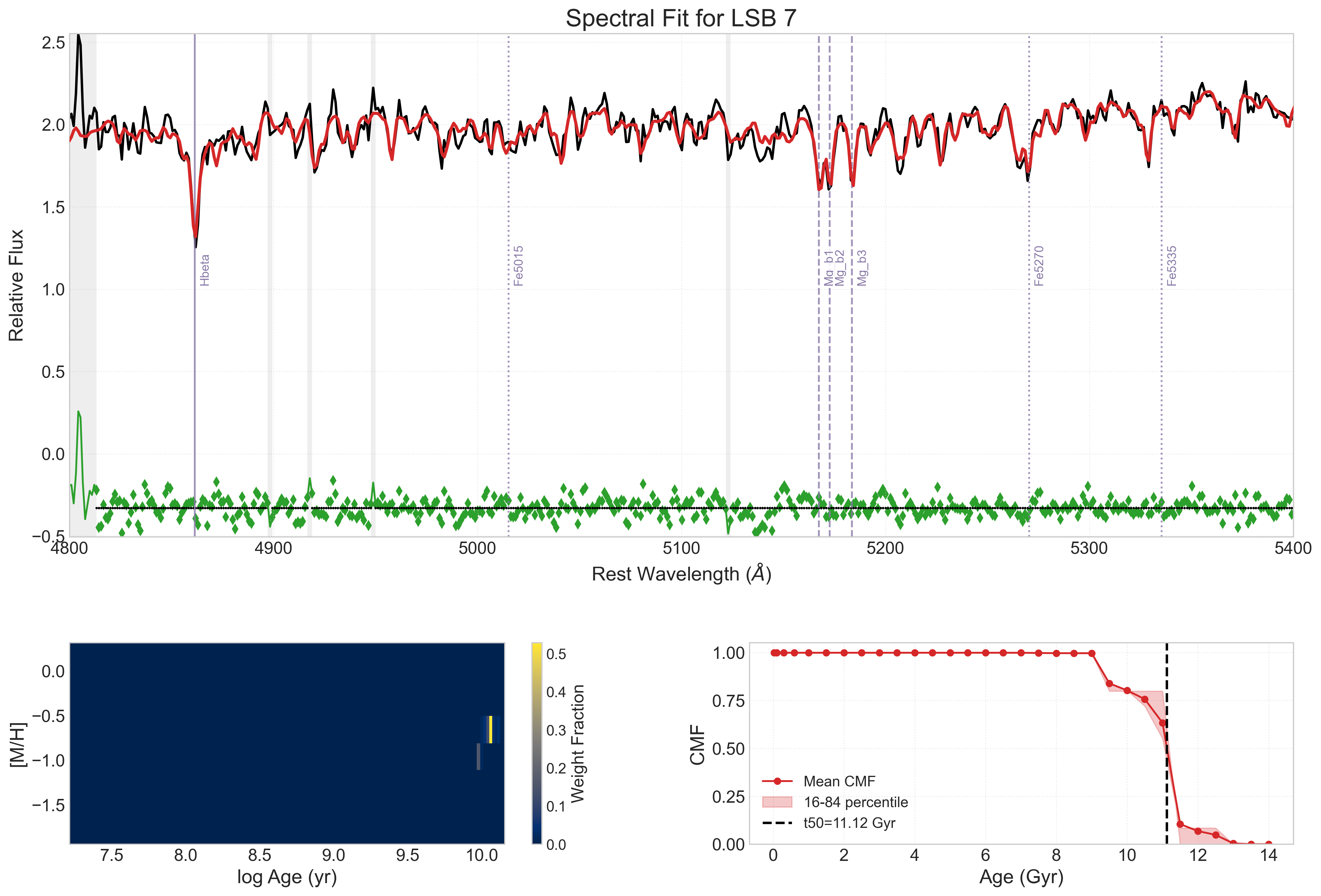}
\caption{Same as Fig.~\ref{fig.UDG1_panels}, but for LSB7.}
\label{fig.LSB7_panels}
\end{figure*}

\begin{figure*}[t]
\centering
\includegraphics[scale=0.44, clip]{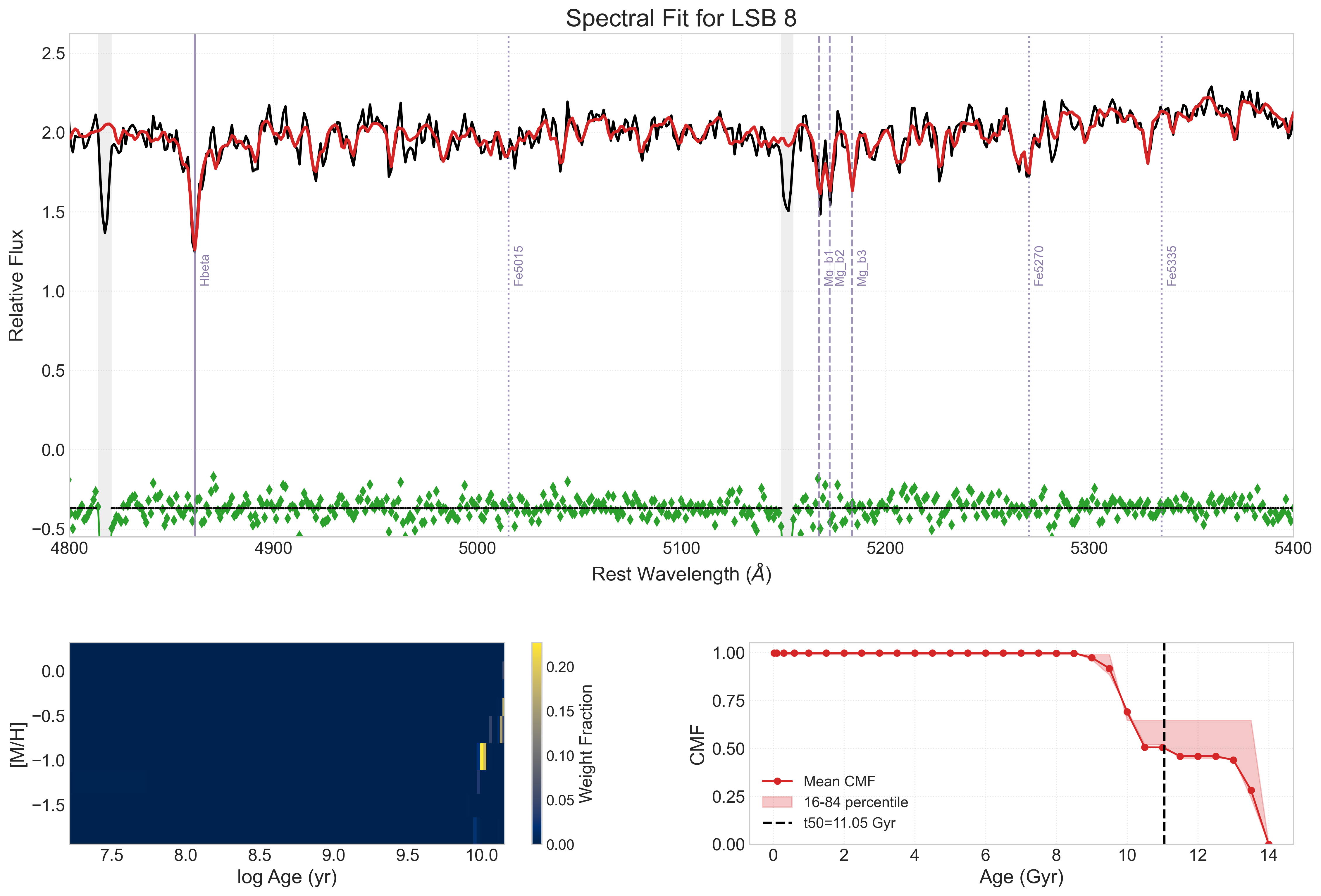}
\caption{Same as Fig.~\ref{fig.UDG1_panels}, but for LSB8.}
\label{fig.LSB8_panels}
\end{figure*}

In this section, we present the stellar population properties and star-formation histories of all the galaxies in LEWIS for which both stellar kinematics and stellar population properties are accurately constrained (see Section~\ref{sec:results}). Figs.\ref{fig.UDG1_panels} to~\ref{fig.LSB8_panels}  reports the {\tt ppxf} fit of the analysed spectral region for the different galaxies, along with a grid showing the median distribution of ages and metallicities obtained through our Monte Carlo procedure. Additionally, we include the cumulative mass fraction obtained by collapsing the age and metallicity weights of each Monte Carlo iteration.

\section{Intermediate cases}
\label{app:intermediate}

In this section, we describe the intermediate cases, i.e., galaxies classified with `I' symbol in Table~\ref{tab:stellar_pop}. These galaxies have spectra with S/N not enough to properly resolve all major absorption features. In addition, the stellar kinematics extraction might be biased and thus affect the reliability of the results from stellar population analysis. These galaxies are UDG3, UDG7, UDG10, LSB1, and LSB4. The spectrum of UDG3 presents a very weak H$\beta$ and lacks any major hint of iron and/or magnesium lines. We thus infer that this source is presumably old, like most of the sample. However, we cannot put a constraint on the exact age or metal content. The same holds for LSB1 and LSB4.
The spectra of UDG7 and UDG10, instead, show clear Fe lines, but the poor quality of the H$\beta$ feature makes the degeneracy of the alpha-enhancement and metallicity hard to break. The large [Mg/Fe] ratio of UDG7 would place this galaxy among the highest alpha-enhanced sources in the sample, as their iron patterns fall outside of solar-scaled model templates. We infer that UDG7 and UDG10 are old and enhanced in magnesium content, yet the large uncertainties and degeneracies due to the data quality make it hard to give significant constraints.

\end{document}